\documentclass[aps,prd,showpacs,amssymb,floatfix,nofootinbib]{revtex4-1}
\usepackage{graphicx,amsmath,subfigure}
\usepackage{epsfig}
\usepackage{dcolumn}
\usepackage{bm}
\usepackage{color}
\usepackage{epstopdf}

\begin{document}
\title{The search for spinning black hole binaries in mock LISA data using a genetic algorithm}
\author{Antoine Petiteau, Yu Shang  and Stanislav Babak}
\affiliation{Max-Planck-Institut fuer Gravitationsphysik,
Albert-Einstein-Institut, \\
Am Muchlenberg 1, D-14476 Golm bei Potsdam, Germany}
\email{Antoine.Petiteau@aei.mpg.de, shangyu@aei.mpg.de, stba@aei.mpg.de, F.Feroz@mrao.cam.ac.uk}
\author{Farhan Feroz}
\affiliation{Astrophysics Group, Cavendish Laboratory, JJ~Thomson Avenue, Cambridge CB3 0HE, UK}
\today

\begin{abstract} 

Coalescing massive Black Hole binaries are the strongest and probably the most
important gra\-vi\-ta\-tio\-nal wave sources in the LISA band. The spin and
orbital precessions bring complexity in the waveform and make the likelihood
surface richer in structure as compared to the non-spinning case. 
We introduce an extended multimodal genetic algorithm which utilizes the properties of the signal
and the detector response function  to analyze the data from the
third round of mock LISA data challenge (MLDC 3.2).  The performance of this
method is comparable, if not better, to already existing
algorithms. We have found all five sources present in MLDC 3.2
and recovered the coalescence time, chirp mass, mass ratio and sky location with reasonable accuracy. 
As for the orbital angular momentum and two spins of the Black Holes, we have found
a large number of widely separated modes in the parameter space with similar maximum likelihood values.


\end{abstract}

\maketitle

\section{Introduction}
\label{S:Intro}

The existence of massive black holes (MBH) with masses ranging from $10^{5} M_{\odot}$ to $10^{9} M_{\odot}$ in 
majority of galactic centers is confirmed by several observations (see  \cite{Ferrarese2005} and reference therein) with  the  MBH in our Galaxy, SgrA* \cite{MBHGalCen_Schodel2002} being the best example. 
Mergers of galaxies are common events in the Universe, it is believed that each galaxy has had at least one merger event during its life. 
During these mergers, the MBH of each galaxy is driven to the center of the remnant full of stars and gas
  via dynamical friction \cite{Callegari2009}. 
The pairs of MBH separated by about kiloparsec are observed in some active galactic nuclei such as  NGC6240~\cite{Komossa2003_NGC6240}, Arp299~\cite{Ballo2004_Arp299}, ESO509-IG066~\cite{Guainazzi2004_ESO509-IG066}, Mrk463~\cite{Bianchi2008_Mrk463} and J100043.15+020637.2~\cite{Comerford2009_J100043}.
The interaction with the gas disc can bring the binary on a tighter orbit down to a few parsecs in a reasonable amount of time (few Myr)~\cite{Dotti2009_DualBH_I}.  There are few candidates of MBH binaries on the sub-parsec scale:
the quasars OJ287~\cite{Valtonen2008_OJ287Nature} ($\sim$ 0.05 pc) and SDSSJ092712.65+294344.0 \cite{Bogdanovic2008_SDSSObs,Dotti2008_SDSSObs}($\sim$ 0.1-0.3 pc). 
To overcome the last parsec separating the MBHs and bring them to the efficient gravitational wave (GW) driven inspiral several scenario have been proposed. 
Here are few possibilities: rotation of the merging galaxies and triaxial potential 
\cite{Berentzen2008AN},   processes involving gas~\cite{Cuadra2009_subpc}, resonant relaxation~\cite{Hopman2006_ResonantRelax}, massive perturber~\cite{Perets2007_MassivePerturber}, young compact stars cluster~\cite{McMillan2003_StarClusterGC}, effect from IMBH~\cite{PortegiesZwart2005_IMBHGalCenter},
etc. When the separation is less than $10^{-3}$ pc, the binary evolution is efficiently driven by the gravitational radiation and can reach the coalescence in less than $10^{9}$ years.
The GWs emitted by  the binary at the end of the  inspiral phase followed by the merger and the ringdown 
 will be detected by the future space born mission LISA with a very high signal-to-noise ratio (SNR). The MBH binaries are the strongest source for LISA and  several such events per year are expected~\cite{Sesana2005_ContributionMBH, Sesana2007_ImprintMBH}.
 It is believed that almost all the MBHs are spinning.  However the predictions for the  magnitudes and directions
 of the spins of MBHs in the binary systems differ  largely depending on the models, the environment of the binary and the physical processes involved (coherent accretion, alignment of spins with the gas disc~\cite{Bogdanovic2007_SpinAlign,  Perego2009_SpinMass, Dotti2009_Spin}, sequence of randomly oriented accretion events~\cite{King2008_Spin}, etc).  Several techniques to measure the spins 
 using electromagnetic radiation~\cite{Brenneman2006_SpinX,Murphy2009_SpinMesHotSpot} have been 
 proposed to measure the spins of MBH binaries, but uncertainties are 
 still very large. The  gravitational wave observations of MBH binaries with LISA should enable us to measure masses and spins of MBHs in the binary with unprecedented accuracy~\cite{SpinBBHLangHugues2006}.
 The knowledge of spins could give us a lot of information about the kick velocity of remnant MBH, the engines 
 of active galactic nuclei, the mechanisms involved in galactic centers, etc. 
 Finally, the spin measurements combined with a precise estimation of masses and sky position made with LISA will increase our understanding of the origin of MBHs, the galactic evolution, the galactic center, cosmology, etc.

Several algorithms for detecting non-spinning MBH binaries in simulated LISA data have already been demonstrated \cite{Brown:2007se, Cornish:2006ms, FerozMN_2009, Gair2009_CS}. 
In this paper we consider inspiralling spinning MBH binaries and we present a particular adaptation of the Genetic Algorithm (GA) to search for GW signals from those systems.  
Genetic algorithms  belong to the family of optimization methods, i.e. they look for extrema. 
The first application of GA in LISA data analysis was proposed in \cite{Crowder2006_GA}  for Galactic binaries. 
In this method the waveform template is associated with an organism, and parameters play the role of the set of genes defining this organism. 
The logarithm of likelihood obtained with a given template defines the quality of the organism. 
A set (colony) of organisms is then evolved through breeding, mutation and custom designed accelerators with the aim of finding the genotype with the highest quality. 
This corresponds to the standard Darwin's principle: weak perishes, strong survives, or, translated into the conventional data analysis language: by evolving a set of templates, we are searching for the parameter set that maximizes the likelihood.

We have applied the GA to the analysis of the third round of mock
LISA data challenge. The mock data set 3.2 consisted of the Gaussian
instrumental noise, Galactic background and between four to six signals from
the inspiralling spinning MBH binaries in a quasi-circular orbit
\cite{Babak:2008sn}. 
We have found several almost equal in value maxima of the likelihood 
which are widely separated in the parameter space. We search for each such strong 
maximum, which we call {\it mode}, and then explore it by a designated set of organisms.
We refer to this extension of the standard GA as a multimodal GA. 
The mutlimodal GA applied to the blind search has shown an excellent
performance: we have detected all present signals with a very accurate
estimation of the parameters (were possible). 

The structure of this paper is as follows. 
In the next Section~\ref{S:Formul} we describe a model of the signal and the 
search template we utilized. Then we discuss the detection statistics we 
 associate with the quality of the organism and their efficient maximization over a subset of 
parameters. In Section~\ref{S:GA} we introduce a Genetic Algorithm and its
particular implementation in GW data analysis. Then, in the Section~\ref{S:Acc}
we introduce accelerators for the rapid evolution -- mechanisms which allow efficiently
explore the parameter space. Besides few standard accelerators often used
in GA we have introduced few new ones, specific to the MBH binary search problem. 
We introduce
multimodal GA (MGA) in Section~\ref{S:Multimode} and describe how 
to identify the modes and to follow their evolution. Our complete algorithm as part of the search 
pipeline is presented 
 in Section~\ref{S:Pipeline} which describes the 
search pipeline. The results are presented and discussed in the Section~\ref{S:Results}
and finally we give a short summary in the concluding Section~\ref{S:Summary}.

\section{Formulation of the problem}
\label{S:Formul}

The mock LISA data challenges are organized in order to encourage the
development of efficient algorithms for gravitational wave data analysis and to 
evaluate their performance.
The third round of MLDC consisted of five challenges but in this work we focus our attention here on 
the data set 3.2 which contained GW signals from 4-6 binaries of spinning MBHs 
(exact number was not revealed to the participants), on top of the confusion Galactic binaries background and 
the instrumental noise. These data was an improvement upon the MLDC challenges 1.2 and 2.2 by adding 
spins to MBHs. The spin-spin and spin-orbital coupling causes the orbital and spins precession which 
results in the modulation of the amplitude and phase of the GW signal. 
The prior range on the parameters and the detailed set up of the challenge can be found in \cite{Babak:2008sn}.


\subsection{Model of the template}
\label{SS:Model}

The signal used in MLDC is modeled as the amplitude-restricted waveform (i.e. only dominant harmonic
at the leading order is used) with the  phase  taken up to the  second Post-Newtonian (PN) order with the leading 
order contributions from the spin-orbital and spin-spin coupling. The binary evolution is described as a
quasi-circular adiabatic inspiral.

The waveform is described by fifteen parameters which are:
the two masses $m_{1}$ and $m_{2}$, 
the time at coalescence $t_{c}$, 
the sky location of the source in ecliptic coordinates, latitude $\beta$ and  longitude $\lambda$, 
the dimensionless spin parameters, $\chi_{1}$ and $\chi_{2}$, 
the initial direction of the spins, polar angles $\theta_{S_{1}}$ and $\theta_{S_{2}}$ and azimuthal angles $\phi_{S_{1}}$ and $\phi_{S_{2}}$, 
the initial direction of the orbital angular momentum, polar angle $\theta_{L}$ and azimuthal angle $\phi_{L}$, 
the phase at coalescence $\Phi_{c}$, 
and the luminosity distance $D_{L}$.

We denote the unit vector in the direction of the orbital angular
momentum as  $\hat{L}$ and two spins are
$\vec{S_1}=\chi_1m_1^2\hat{S}_1$, $\vec{S_2}=\chi_2m_2^2\hat{S}_2$, where $\hat{S}_{1,2}$ are unit
vectors and $0 < \chi_{1,2} < 1$. The precession equations are given
in \cite{ACST_spin}

\begin{eqnarray}
\dot{\vec{S_1}}=\frac{(M\omega)^2}{2M}\left\{\eta(M\omega)^{-1/3}\left(4+\frac{3m_2}{m_1}\right)\hat{L}
+\frac{1}{M^2}\left[\vec{S_2}-3(\vec{S_2}\cdot \hat{L})\hat{L}\right]\right\}\times\vec{S_1}
\label{eq:S1},
\end{eqnarray}

\begin{eqnarray}
\dot{\vec{S_2}}=\frac{(M\omega)^2}{2M}\left\{\eta(M\omega)^{-1/3}\left(4+\frac{3m_1}{m_2}\right)\hat{L}
+\frac{1}{M^2}\left[\vec{S_1}-3(\vec{S_1}\cdot \hat{L})\hat{L}\right]\right\}\times\vec{S_2}
\label{eq:S2},
\end{eqnarray}

\begin{eqnarray}
\dot{\hat{L}}=-\frac{(M\omega)^{1/3}}{\eta M^2} \left( \dot{\vec{S_{1}}}+ \dot{\vec{S_{2}}} \right)
&=&\frac{\omega^2}{2M} \Bigg\{\left[\left(4+\frac{3m_2}{m_1}\right)\vec{S_1}+\left(4+\frac{3m_1}{m_2}\right)\vec{S_2}\right]\times\hat{L}\nonumber\\
&&-\frac{3\omega^{1/3}} {\eta M^{5/3}}
\left[\left(\vec{S_2}\cdot\hat{L}\right)\vec{S_1}+\left(\vec{S_1}\cdot\hat{L}\right)\vec{S_2}\right]\times\hat{L}
\Bigg\}.
\label{eq:L}
\end{eqnarray}
The modulation of the waveform  due to the presence of spins is taken at the leading order.

The orbital angular frequency with spin effect up to 2~PN order is given by
\begin{eqnarray}
M\omega &=&
\frac{1}{8}\tau^{-3/8}\Bigg[1+\left(\frac{743}{2688}+\frac{11}{32}\eta\right)\tau^{-1/4}-\frac{3}{10}\left(\pi-\frac{\beta}{4}\right)\tau^{-3/8}\nonumber\\
&&+\left(\frac{1855099}{14450688}+\frac{56975}{258048}\eta+\frac{371}{2048}\eta^2-\frac{3}{64}\sigma\right)\tau^{-1/2}\Bigg],
\end{eqnarray}
where  $M=m_1+m_2$ is total mass, $\eta=\frac{m_1 m_2}{M^2}$ is the symmetric mass ratio and
\begin{eqnarray}
\tau&=&\frac{\eta}{5M}(t_c-t)\\
\beta&=&\frac{1}{12}\sum_{i=1,2} \left[ \chi_i \left( \hat{L} \cdot \hat{S_i} \right) \left( 113\frac{m_i^2}{M^2}+75\eta \right) \right] \\
\sigma&=&-\frac{1}{48} \eta \chi_1 \chi_2 \left[ 247 \left( \hat{S_1}\cdot\hat{S_2} \right) - 721 \left( \hat{L} \cdot \hat{S_1} \right) \left( \hat{L} \cdot\hat{S_2} \right) \right].
\end{eqnarray}
In our following analysis, we use $\eta$ and the chirp mass $M_c=\frac{(m_1m_2)^{3/5}}{(m_1+m_2)^{1/5}}$ 
as independent parameters  instead of $m_{1}$ and $m_{2}$.


The intrinsic phase is 

\begin{eqnarray}
\Phi_{\textmd{orb}}&=&\Phi_C-\frac{\tau^{5/8}}{\eta}\Bigg[1+\left(\frac{3715}{8064}+\frac{55}{96}\eta\right)\tau^{-1/4}-\frac{3}{16}\left(4\pi-\beta\right)\tau^{-3/8}\nonumber\\
&&+\left(\frac{9275495}{14450688}+\frac{284875}{258048}\eta+\frac{1855}{2048}\eta^2-\frac{15}{64} \sigma\right)\tau^{-1/2}\Bigg].
\end{eqnarray}
%
The phase is defined with respect to the ascending node, however the spin-orbital coupling causes 
precession of the orbit, therefore we need to introduce precessional correction to the phase, $\delta\Phi(t)$.
It depends on the unit vector $\hat{n}$ pointing from the solar system barycenter to the source: 
\begin{eqnarray}
\Phi(t)&=&\Phi_{\textmd{orb}}(t)+\delta\Phi(t),\\
\dot{\Phi}(t)&=&\omega+\delta\dot{\Phi}=\omega+\frac{(\hat{L}\cdot\hat{n})(\hat{L}\times\hat{n})\cdot\dot{\hat{L}}}{1-(\hat{L}\cdot\hat{n})^2},\\
\delta\Phi(t)&=&-\int^{t_c}_t\left(\frac{\hat{L}\cdot\hat{n}}{1-(\hat{L}\cdot\hat{n})}\right)(\hat{L}\times\hat{n})\cdot\dot{\hat{L}}dt.
\end{eqnarray} 

The gravitational wave polarization components in the source frame are given by 

\begin{eqnarray}
h_+&=&  h_{+ 0} \cos2\Phi = -\frac{2M\eta}{D_L}(1+\cos^2\iota)(M\omega)^{2/3}\cos2\Phi \nonumber \\
h_{\times}&=& h_{\times 0} \sin2\Phi = \frac{4M\eta}{D_L}\cos\iota(M\omega)^{2/3}\sin2\Phi ,
\label{E:hpc}
\end{eqnarray}
where $\cos\iota=\hat{L}\cdot\hat{n}$.

The strain $h(t)$ that the GW exerts on the LISA detector is the
following linear combination of $h_+$ and $h_{\times}$
\begin{eqnarray}
h(t)=F_+(\beta,\lambda)h^S_+(t)+F_{\times}(\beta,\lambda)h^S_{\times}(t),
\label{E:hGlobal}
\end{eqnarray}
where $F_+$ and $F_{\times}$ are ``detector beam-pattern'' coefficients.
The polarization components in the radiation frame, $h^S_+$ and $h^S_{\times}$, are expressed as
\begin{eqnarray}
h^S_+&=&-h_+\cos2\psi-h_{\times}\sin2\psi,
\nonumber \\
h^S_{\times}&=&h_+\sin2\psi-h_{\times}\cos2\psi,
\label{E:hSpc}
\end{eqnarray}
where the polarization angle $\psi$ is defined by
\begin{eqnarray}
\tan\psi=\frac{\sin\beta\cos(\lambda-\phi_L)\sin\theta_L-\cos\theta_L\cos\beta}{\sin\theta
_L\sin(\lambda-\phi_L)}.
\label{E:Psi}
\end{eqnarray}
Note, due to the precession of the orbital plane, this polarization angle varies during the evolution of the binary.

The data sets in MLDC are the TDI (time delay interferometry) variables. These are the combinations of  the time
delayed measurements which drastically reduce the  laser frequency noise \cite{VallisCsqPerteLiens, lrr-2005-4}.
In our search, we adopted the two orthogonal (i.e. with uncorrelated noise) TDI  channels, A and E, in the phase domain (i.e. strain).
In our template, we consider a long wavelength approximation to these signals \cite{Cornish:2002rt, ThesePetiteau}.
This approximation ($L\omega \ll 1$, where $L$ is LISA' armlength  and $\omega$ is an instantaneous frequency of GW)
 works pretty well below approximately 5 mHz. Assuming rigid LISA with equal arms, the waveform sampled at discrete times takes
 the following form \cite{Cornish:2002rt, ThesePetiteau}

\begin{eqnarray}
 h_{I} (t_k) &\simeq & 2 L \ \sin \Delta \phi_{2L}(t_{k}) \times \nonumber\\ 
  & & \left\{ - h_{+0}(t_{k}) \left[ \cos{(2 \psi(t_{k}) )}  F_{+I}(t) - \sin{(2\psi(t_{k}) )}  F_{\times I}(t)  \right] \sin \phi' (t_{k})  \right.  \nonumber \\
& & \left.  + h_{ \times 0}(t_{k})  \left[ \sin{(2 \psi(t_{k})  )}  F_{+I}(t)  + \cos {(2 \psi(t_{k})  )}  F_{\times I}(t)  \right]  \cos \phi' (t_{k})  \right\},
\label{E:TDILW}
\end{eqnarray}

where $I=\{ A, E \}$,  $ \Delta
\phi_{2L} (t) = ( \phi (t_{k}) - \phi (t_{k} - 2L) ) / 2 $, 
$ \phi'(t) = ( \phi (t_{k}) + \phi (t_{k} - 2L) ) / 2 $ with $\phi(t)$
being the phase of GW and
$t_{k} = t + \widehat{n}.\overrightarrow{r_{0}} $ is the time in LISA frame with $\overrightarrow{r_{0}}$ the vector from the Sun to LISA barycenter.
The antenna pattern functions $F_{+I}$ and
$F_{\times I}$ corresponding to the TDI channel, have the following expressions 

\begin{eqnarray}
F_{+} (\theta_{d},\lambda_{d}; t, \Omega ) & =  & {1 \over 32} \left[ 6 \sin ( 2\theta_{d} ) \left( 3 \sin\left( \Phi_{T}(t) +
\lambda_{d} +  \Omega \right) -  \sin\left( 3  \Phi_{T}(t) - \lambda_{d} +  \Omega \right) \right) \right. \nonumber \\
 & &  \left. - 18 \sqrt{3} \sin^{2} \theta_{d} \sin \left( 2 \Phi_{T}(t) + \Omega \right) \right.
 - \sqrt{3} \left( 1 + \cos^{2}\theta_{d} \right) \times
 \nonumber  \\
&  &  \left. \left( \sin \left(4  \Phi_{T}(t) - 2 \lambda_{d} + \Omega \right) + 9 \sin \left(2\lambda_{d} + \Omega \right) \right) \right] \label{E:TDILW_Fp}, \\
F_{\times} (\theta_{d},\lambda_{d}; t, \Omega ) & =  & {1 \over 16} \left[ \sqrt{3} \cos \theta_{d} ( \cos ( 4  \Phi_{T}(t) - 2 \lambda_{d} + \Omega ) -  \right. \nonumber \\
&  & \left.   9 \cos ( 2 \lambda_{d} + \Omega ) )+  6 \sin \theta_{d} ( \cos ( 3  \Phi_{T}(t) - \lambda_{d} + \Omega ) +  \right. \nonumber  \\
& & \left.  3 \cos (  \Phi_{T}(t) + \lambda_{d} + \Omega ) ) \right]
\label{E:TDILW_Fc}
\end{eqnarray}
with $\theta_{d} = \beta + \pi/2 $, $\lambda_{d} = \lambda + \pi$,
$\Phi_{T}(t) = 2 \pi t / Year$ and  $\Omega = 0, -\pi/2$ for channels A and E respectively. Note that this is the long wavelength approximation to the 
signal injected in the simulate data, we found it to be a reasonably accurate representation 
until the last $1-1/2$ cycles before the merger. The end of the signal is discussed in more detail later.

\subsection{quality estimation}
\label{SS:quality}

The signal from one detector is
\begin{equation}
s_i(t)=h_i(t,\hat{\theta})+n_i(t),
\label{E:SigEqHpN}
\end{equation}
where $h_i(t,\hat{\theta})$ is a signal described by a set of
parameters $\hat{\theta}$ and $n_i(t)$ is the stationary Gaussian
noise characterized by the power spectral density (PSD) $S_n(f)$, $$S_n(f)\delta(f-f')
=2 \overline{\tilde{n}(f) \tilde{n}(f')}.$$ Here the over-bar means the average over ensemble 
of the noise realizations and the tilde denotes the Fourier transform defined as
\begin{equation}
\widetilde{n}(f) = \int_{-\infty}^{\infty} dt \  n(t) e^{2 \pi i f}.
\label{E:FourierTransform}
\end{equation}

We use the maximum likelihood method (ML) \cite{
MFFinn, Owen1996, Cutler:1994ys}
for estimating the parameters of the waveform
$\widehat{\theta}$.  Assuming that the noise $n(t)$ is a normal
process with zero mean, then the likelihood (probability) of the presence of signal $h(\hat{\theta})$
in the detector output  is given by 
\begin{equation}
p (s|\hat{\theta}) = C e^{- \langle s - h(\hat{\theta}) \mid s -
h(\hat{\theta}) \rangle /2},
\label{E:Likelihood}
\end{equation}
where $\langle h \mid s \rangle$ is the inner product of the signal
given by
\begin{equation}
\langle h \mid s \rangle  =  2 \int_{0}^{\infty} df \
{\widetilde{h} (f) \; \widetilde{s}^{*} (f) + \widetilde{h}^{*} (f)
\; \widetilde{s} (f) \over S_{n}(f)  } .
\label{E:InnerProduct}
\end{equation}

The noise is the sum of instrumental noise
$S^{\textmd{inst.}}_{n} (f)$ and the GW confusion noise from Galactic binaries 
$S_{n}^{\textmd{Gal. Bin.}} (f)$. In strain data (i.e. phase measurements),
the instrumental noise for TDI variables A and E is described by the following PSD
\begin{eqnarray}
S^{\textmd{inst.}}_{n} (f) & = &    16 \sin^{2}( \phi_{L}) \left( S_{n}^{\textmd{OPN}} (f)  + \left(  \cos (2 \phi_{L}) \right) S_{n}^{\textmd{acc.}} (f) \right) \nonumber \\
& & - 4 \sin( 2 \phi_{L})  \sin ( \phi_{L})  \left( 4
S_{n}^{\textmd{OPN}} (f)  + S_{n}^{\textmd{acc.}} (f) \right),
\label{E:NoiseTDI}
\end{eqnarray}
where the acceleration noise is 
$ S^{\textmd{acc.}}_{n} (f) = 5.75 \times 10^{-53} ( f^{-4} + 10^{-8}  f^{-6} ) \ Hz^{-1} $
 and the optical path noise and the shot noise are 
 $ S^{\textmd{OPN}}_{n} (f) =  3.675 \times 10^{-42} \ Hz^{-1} $.

The Galactic GW confusion noise is a combination of the unresolved signals from $\sim 30$ 
millions of white dwarf binaries. This noise is modeled by the following function, in units  
of $Hz^{-1},$ \cite{Nelemans2004_GalWD, TimpanoCornishRubbo}
\begin{equation}
S_{n}^{\textmd{Gal. Bin.}} (f) = \left\{
\begin{array}{ll}
10^{-44.62}  f^{-2.3} &  10^{-4} \leq f \leq 10^{-3} \\
10^{-50.92}  f^{-4.4} &  10^{-3} \leq f \leq 10^{-2.7} \\
10^{-62.8}  f^{-8.8} &  10^{-2.7} \leq f \leq 10^{-2.4} \\
10^{-89.68}  f^{-20} &  10^{-2.4} \leq f \leq 10^{-2}
\end{array}
\right. \label{E:NoiseGalBin}
\end{equation}


\subsubsection{Maximization of the likelihood}
\label{SS:MaxL}

The goal of the method presented in this paper is to find the maximum of the
likelihood in the 15-dimensional parameter space, and, thus, obtain the maximum
likelihood estimation of the parameters of the signal. The value of the likelihood 
tells us also about the statistical significance of the detected event. In case of LISA 
data, the signals usually have high signal-to-noise ratio (SNR), so the probability 
of the false detection is rather low. However, the data is signal dominated and 
several GW signals of one type (say, Galactic binaries) could conspire and produce
significantly high SNR at the output of the matched filtering during the search 
for another type of signal (say, SMBH binary) \cite{Racine2007_Gaussinity}.


It is possible to maximize the likelihood analytically over two parameters and we will call the 
resulting  function  Maximized Likelihood (or quality). The procedure of maximization is similar
 to the one used to produce the $\mathcal{F}\textrm{-statistic}$ \cite{Jaranowski:1998qm, Cornish:2006ms}. 
 Due to amplitude modulation we can only maximize over two parameters: the luminosity distance $D_{L}$
and the phase at coalescence $\Phi_{c}$.

We will be working with the logarithm of the likelihood (getting rid of the normalization factors which does not depend on
parameters by adopting the relative likelihood):

\begin{equation}
\ln \mathcal{L} \simeq \sum_I  \langle s_I \mid h_I \rangle - {1 \over 2} \sum_I  \langle h_I \mid h_I \rangle ,
\label{E:Fs_lnL1}
\end{equation}
here the sum is over the independent detectors (TDI streams, $I=\{A, E\}$).
The GW template \eqref{E:TDILW} can be express as
\begin{equation}
h_I = \sum_k a_k \ h_{kI} 
\label{E:Fs_hI}
\end{equation}
and the extrema of $\ln \mathcal{L}$ over $a_{k}$ are found by solving
the coupled set of equations
\begin{eqnarray}
\frac{\partial\ln \mathcal{L}}{\partial a_k}=0
\end{eqnarray}
and it turns out that
\begin{equation}
\quad a_k = X_k (M_{kl})^{-1} \quad \quad \textrm{with} \; X_{k} =
\sum_{I} \langle s | h_{kI} \rangle,  \;\; M_{kl} = \sum_{I} \langle
h_{kI} | h_{lI} \rangle .
\label{E:Fs_ak}
\end{equation}
The log likelihood maximized over $a_k$ is called the $\mathcal{F}$-statistic in the case where $a_k$ are four functions of (constant)
amplitude, polarization angle, inclination and initial phase:
\begin{equation}
\mathcal{F} \simeq  {1 \over 2} X_k (M_{kl})^{-1} X_l.
\label{E:Fs_XM}
\end{equation}

In the case of spinning MBH binary, the analytic maximization is possible only over the luminosity distance 
$D_{L}$ and the phase at coalescence $\phi_{c}$. Consequently, the dimension of the search
parameter space is reduced to 13. Following \eqref{E:Fs_hI} the template \eqref{E:TDILW} 
 can be expressed as
\begin{equation}
 h_{I} (t) = a_1 \ h_{1I}(t) + a_2 \ h_{2I}(t)
\label{E:Fs_hIa1a2}
\end{equation}
 with
\begin{equation}
\left\{ \begin{array}{l}
a_1 = \cos (2 \phi_c) / D_L \\
a_2 = \sin (2 \phi_c) / D_L
\end{array} \right. , \;\;\;
\left\{ \begin{array}{lll}
h_{1I} = h_I (D_L = 1 Gpc ; \ \phi_c = 0 ) \\
h_{2I} = h_I (D_L = 1 Gpc ; \ \phi_c = \pi/4 )
\end{array} \right.
\label{E:Fs_h1h2}
\end{equation}

Using the above expressions and the orthogonality 
$\tilde{h}_{2I} \simeq i\ \tilde{h}_{1I} $ we obtain the maximized 
log likelihood.

\begin{equation}
\mathcal{F} \simeq { {\left( \sum_I <s_I |h_{1I}> \right)}^2 + {\left(
\sum_I <s_I |h_{2I}> \right)}^2 \over {\left( \sum_I <h_{1I}|h_{1I}>
\right)}^2 } \label{E:Fs_Approx}
\end{equation}

\subsubsection{Maximization over the time of coalescence}
\label{SS:MaxTc}
In order to efficiently find the time of coalescence, we use 
correlation in place of the inner products. Given a  template $h$ which is 
constructed with the initial value (usually taken at the edge of the prior) 
$t_{c,0}$ and using the inverse Fourier transform, we find the value of $\tau_{\textmd{max}}$ 
which maximizes \eqref{E:Fs_Approx} or which is equivalent to maximizing
\begin{equation}
c(\tau)  =  2 \int_{0}^{\infty} df \  {\widetilde{h} (f) \;
\widetilde{s}^{*} (f) + \widetilde{h}^{*} (f) \; \widetilde{s} (f)
\over Sn(f)  } e^{- 2 i \pi f \tau} \label{E:TcMax}.
\end{equation}
Note that the amplitude of the signal depends on the choice of $t_c$ via annual modulation caused 
by LISA's orbital motion, therefore the new value $t_{c,1}=t_{c,0}+\tau_{\textmd{max}}$
is not necessarily the final answer. The time of coalescence  which maximizes the quality 
\eqref{E:Fs_Approx} for given other parameters should correspond to maximum of \eqref{E:TcMax}
at zero (or almost zero) lag. Using the new value of $t_c$ we repeat the maximization, and we stop
iterations when the difference $\left|t_{c,i} - t_{c,i-1} \right|$ is sufficiently small. 
Usually few iterations are sufficient to find $t_c$ which maximizes the quality.  


\subsubsection{The waveform termination}
\label{SS:CutWaveform}

 The signal from MBH binaries
is band limited, the lower frequency limit is defined approximately by twice the orbital phase at $t=0$.

The upper frequency is introduced somewhat arbitrarily. To terminate both the signal and the template
smoothly an exponential taper  is applied.  The taper affects the data when  two black holes 
are separated by a distance $R = 7 M$  and kills the signal completely around $R=6M$  
(which is the last stable orbit for the test mass in Schwarzschild space-time). 
Therefore, in computing the overlaps, we use the maximum frequency in the integration corresponding 
to the orbital separation $6M$: 
\begin{eqnarray}
f_{\textmd{max}}=\frac{1}{\pi M(R/M)^{3/2}}=\frac{\eta^{3/5}}{\pi
(R/M)^{3/2}M_c}.
\label{E:fmax}
\end{eqnarray}

The exponential taper causes problems for the long-wavelength approximation, and  our template
deviates from the signal during the last cycle. Unfortunately these small deviations fall in the 
most sensitive part of the LISA band and are further enhanced by high SNR.  This causes a 
significant problem:
the bias caused by this deviation is unacceptably large because there is a large region of the parameter 
space that produces templates which fit the end part of the signal perfectly (using incorrect parameters) but
fail to reproduce the low frequency part of the signal.


In order to solve this problem we terminate the template waveform  few cycles earlier by fixing cutoff 
frequency which corresponds to the orbital separation $R>7M$. Our approximation becomes better 
as we go to lower frequencies, however we start losing power of the signal (SNR) which is highly 
undesirable.  We automatically readjust the frequency cut-off if the SNR drops below a certain threshold. 



\begin{figure}[!htbp]
\center
\includegraphics[width=0.45\textwidth, clip=true, viewport=0 15 269 150]{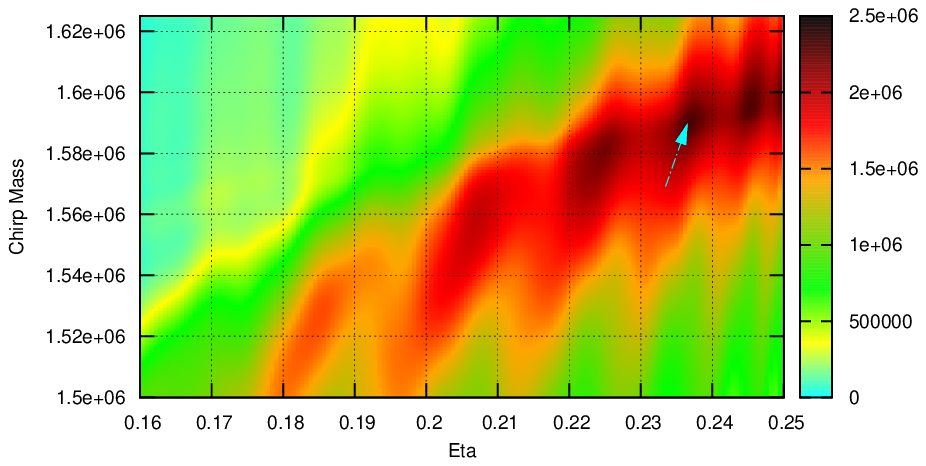}
\includegraphics[width=0.45\textwidth, clip=true, viewport=0 15 269 150]{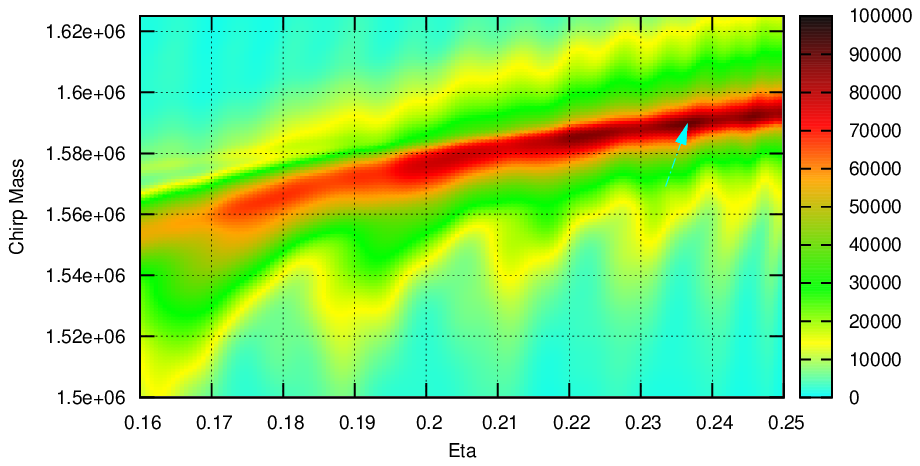}
\caption{Distribution over $M_{c}$ and $\eta$, of the Maximized Likelihood (quality) computed with the full waveform on left panel and with the waveform cut at $f_{max} = 0.26 \ \textrm{}$ on the right panel. This example corresponds to a signal with the following parameters: 
$\beta = -0.38896 \ rad$, 
$\lambda = 3.28992\ rad$, 
$ t_{c} =  19706568.3273 \ sec$, 
$M_{c} =  1589213.34 \ M_{\odot}$,  
$\eta = 0.23647$,
$\theta_{L} = 2.78243\ rad$,
$\phi_{L} =  1.53286\ rad$, 
$\chi_{1} =  0.24115$, 
$\chi_{2}  =  0.16145$,  
$\theta_{S1} = 1.20839\ rad$, 
$\phi_{S1} = 5.61808\ rad$, 
$\theta_{S2} = 0.39487\ rad$, 
$\phi_{S2} = 5.82937\ rad$, 
$D_{L} = 6856164697.8 \ parsec$, 
$\phi_{c} = 4.96746 \ rad$
. The arrow points to the true parameters.}
\label{F:DistFMcEtaCutFull}
\end{figure}

We want to emphasize a very important feature which accompany the earlier termination of the waveform.
The map of the quality changes: in the Figure~\ref{F:DistFMcEtaCutFull} we show the map of the quality in 
the ``chirp mass'' - ``eta'' plane keeping other parameters fixed to their true values. On the left panel we show 
$\mathcal{F}_{full}$ (we use no frequency cut off other that introduced by the taper), and, on the right 
panel, we plot $\mathcal{F}_{cut}$ with template cut at $f_{max} = f_{cut} = 0.26\, \textmd{mHz}$. One can see multiple maxima in both plots, but(!) the position of the secondary maxima are different
whereas the location of the true (global) maximum (indicated by an arrow) is the same. It can also be seen
 that the size of the secondary maxima on the right panel is smaller. We will use these features later in 
our search. 


\subsubsection{$A$-statistic}
\label{SS:Astatistic}

\begin{figure}[!htbp]
\center
\includegraphics[width=0.45\textwidth, clip=true, viewport=0 15 240 150]{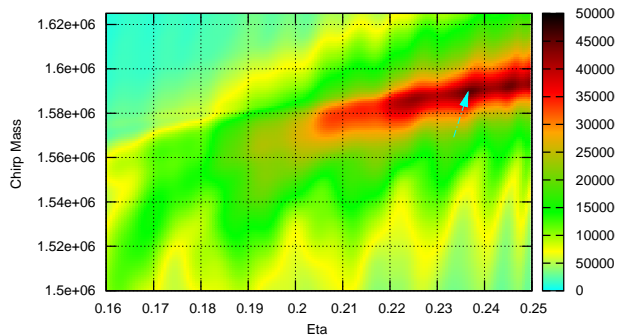}
\caption{Distribution of $A$-statistic over $M_{c}$ and $\eta$. This example corresponds to the same signal as in Figure~\ref{F:DistFMcEtaCutFull}. The arrow points to the true location of parameters of the signal.}
\label{F:DistAMcEta}
\end{figure}

Chopping the template at lower frequency solves the problems mentioned above but is not completely 
satisfactory. We lose some SNR and  consequently some accuracy in the parameter estimation, 
we also lose information stored at the end of the signal which is especially important to recover
spin-related parameters.  In order to reduce the impact of the coalescence part, without killing it completely. For that, we introduce a new function, called {\it $A$-statistic} which is simply a geometrical mean of the Maximized Likelihood of the cut waveform and the Maximized Likelihood of the full waveform:
\begin{eqnarray}
\mathcal{A}=\sqrt{\mathcal{F}_{\textmd{cut}}\times\mathcal{F}_{\textmd{full}}}.
\label{E:Astatistic}
\end{eqnarray}
$A$-statistic is not log likelihood anymore, but one of its advantages is that it keeps 
the information from the full waveform including the coalescence but at the same time it enhances the 
information coming from the low-frequency part. $A$-statistic also reduces the 
number of local maxima as can be seen in the Figure~\ref{F:DistAMcEta}. 
 In this example we have reduced the size and number of maxima from five to three. 
 %

\section{Genetic algorithm}
\label{S:GA}
\subsection{The basic principle}
In order to find all the parameters of the signal, we need an
effective algorithm to search over the 13 dimensional parameter space.
Building the grid in the multi-dimensional parameter space is a highly non-trivial problem.
 The use of the stochastic/random bank \cite{Babak:2008rb, Harry2008, Messenger2009,
 Manca2009} is a feasible method for the template placement, however
 a full grid scan over the whole parameter space would be prohibitively
computationally expensive.
 Alternative would be to use variations of the Markov chain Monte-Carlo
 \cite{Cornish:2006ms} or nested sampling
\cite{FerozMN_2009} methods. Here we have chosen to use genetic 
 algorithm (GA) (adjusted to our needs) to search for the global maximum of the likelihood in multi-dimensional
 parameter space.

The GA is derived from the computer simulations of the biological system,  
which were originally introduced by Professor Holland and his students
in Michigan University.
It is a method for the global search (optimization method) 
 based on the natural selection principle -- the basis for the 
evolution theory established by C. Darwin. In the nature, 
organisms adapt themselves to their environment: the
smartest/strongest/healthiest organisms are more likely to survive
and participate in the breeding to produce the offsprings. 
These two processes, selection and breeding, are used in genetic algorithms
to produce a new generation of organisms.  Since the best organisms are more
likely to participate in breeding, the new generation should be
better than the previous one (at least no worse). So this procedure
induces the evolution of the organism, just like in the nature, the
good qualities of the parents can be transferred to their offsprings. 
In the biological world, besides  these two basic operations, among every generation, 
there are always  few individuals which have better characteristics to adopt to the environment,  produced as a result of a positive mutation. By introducing the new genotype into the 
population, mutation can potentially
improve the forthcoming generations and consequently accelerate the evolution
towards the global maximum. 

Some measure of ``goodness'' needs to be associated with each organism. 
In the case of gravitational wave search, it is natural to associate the logarithm of the likelihood 
(or any other equivalent detection statistic e.g. Maximized Likelihood or $A$-statistic discussed in the
 previous section) with the ``goodness'' which needs to be ``improved'' through the evolution 
 of the organisms. We will call the chosen measure of ``goodnees'', the {\it quality} of an organism $Q$.

Following is a brief description of how a typical GA works. We start with a randomly chosen 
group of organisms (templates), we evaluate the quality of each organism (log likelihood).
We select set of pairs (parents) based on their quality, the organisms with better
quality (templates with higher likelihood) are chosen more often than weak organisms.
We combine genotype of two parents to produce a child (we combine parameters
of two chosen templates to produce a new one). Number of produced children  is equal to the 
number of parents (we keep number of evolving organisms (generation) fixed).
Next we allow  with a certain probability a random mutation in the children's genes
(with some probability we randomly change the parameters of the new templates, 
exploring a larger area of the parameter space). The parents are discarded and the resulting children form a new generation. We repeat 
the procedure until we reach steady state (maximum in the quality). In this simple
example we keep only one generation active (one group of templates).


A list of (biological) GA terms with the equivalent terms in GW data analysis is given in the Table~\ref{T:DicoGWsearchGA}.

\begin{table}[htdp]
\begin{center}
\begin{tabular}{ccc}
Genetic algorithm & & GW search \\
\hline
organism & $\Longleftrightarrow$ & template \\
gene (of an organism) & $\Longleftrightarrow$ & parameter (of a template) \\
allele (of a gene) & $\Longleftrightarrow$ & bits (of the value of the parameter) \\
quality $Q$ & $\Longleftrightarrow$ & Maximized Likelihood or $A$-statistic\\
colony of organisms & $\Longleftrightarrow$ &  evolving group of templates \\
$n$-th generation & $\Longleftrightarrow$ & the state of colony at $n$-th step of evolution \\
(selection + breeding) + mutation & $\Longleftrightarrow$ & way of exploring  the parameter space
\end{tabular}
\end{center}
\caption{Relation between GA  and GW notions.}
\label{T:DicoGWsearchGA}
\end{table}%

In the following subsections we give a detailed description of each element of the basic GA and then  
we introduce  the specific modifications to speed up the search.


\subsection{Code of the gene}
\label{SS:GACodeGene}

As we have discussed above, every organism is associated with a template and the parameters 
of the template play the role of genes. So each organism is described by 15 genes, two of them are 
chosen optimally (maximization of the log likelihood, see Sec.~\ref{SS:MaxL}) and the gene corresponding to the time of 
coalescence is efficiently found using correlation (see Sec.~\ref{SS:MaxTc}). We imitate the DNA structure by describing the gene (parameter value) by a set of alleles. In our implementation we adopt a binary representation of 
the gene (parameter) which means that each allele (bit) has two possible values: 0 or 1. In practice 
we first fix the precision of each parameter (by fixing the number of significant digits in the decimal format)
and then we translate it to standard binary and/or in Gray form. In our method we use both
representations, the reason will be explained later when we discuss quantization issue.

 
Let us show how this is done in practice.
Consider a parameter $\theta_{k}$ with the  uniform prior range $[x_{k,min}, x_{k,max}]$. 
First we convert a value $x_{k}$ of $\theta_{k}$ into an integer $c_{k} = (x_{k} - x_{k,min}) / \Delta x_{k}$ 
where $\Delta x_{k} = (x_{k,max} - x_{k,min}) / 2^{N_{k}}$ is the resolution of $\theta_{k}$ 
and $N_{k}$ is the number of bits.
Then, we convert $c_{k}$ into the set of bits $b_{k}[i]$ using the coding rule of the chosen representation.
As we see, the resolution for each parameter depends on the number of bits $N_{k}$ used 
for describing it and is the same for both representations. The importance of the bit is determined by 
its position. A change of a bit in a higher position (significant bit) corresponds to a big change
in the parameter value. In our convention, the first bit, $b_{k}[0]$, is the lowest significant bit 
and the last bit, $b_{k}[N-1]$, is the highest significant bit.


There is a close relationship between two gene representations.
We can transform the binary representation to Gray representation by the
following procedure: 
given a string of binary code with $N$ bits
$\{B[0],B[1],\cdots,B[N-1]\}$, we set $B[N]=0$ then the Gray code
with the same N bits is
\begin{eqnarray}
G[i]=B[i+1] \wedge B[i],
\end{eqnarray}
where the operator ``$\wedge$'' corresponds to the XOR operator in 
programming languages.
Other way round, by setting again $B[N]=0$, we can  get the
binary representation with $N$ bits from the Gray representation as
\begin{eqnarray}
B[i]=B[i+1] \wedge G[i].
\end{eqnarray}


\subsection{Selection}
\label{SS:Selection}
The selection process chooses the parents for breeding. 
The probability of selecting an organism is defined by its quality. 
Organisms with higher quality have a better chance
of being chosen to participate in the breeding.  
First the quality $Q_{i}$, i.e. the Maximized Likelihood or $A$-statistic, for all organisms is computed
(index $i$ refers to the $i$-th organism). 
Then each organism is assigned the probability of being chosen for breeding as $p_i=Q_{i} / \sum_{j}^{N} Q_{j}$.
The selection is made by the roulette selection method:  
we choose a random number uniformly from [0,1];
if it is bigger than $p_i$ and smaller than $p_{i+1}$, then the $i^{th}$ organism is selected.
This selection ensures that the ``good'' organisms are chosen more often than the ``bad'' ones and
 guarantees that the genotype responsible for a high quality propagates in generations 
approaching the optimal value.


In our approach we do not take into account the geographical proximity
between parents (in other words possible correlation between
templates in the same generation). By forbidding the breeding
between the correlated parents, it might be more efficient to explore a
larger region of parameter space, but the overall resolution of the method will be
reduced. We therefore do the selection based only on the quality.


\subsection{Breeding}
\label{SS:Breeding}
\begin{figure}[!htb]
\center
\includegraphics[width=0.3\textwidth]{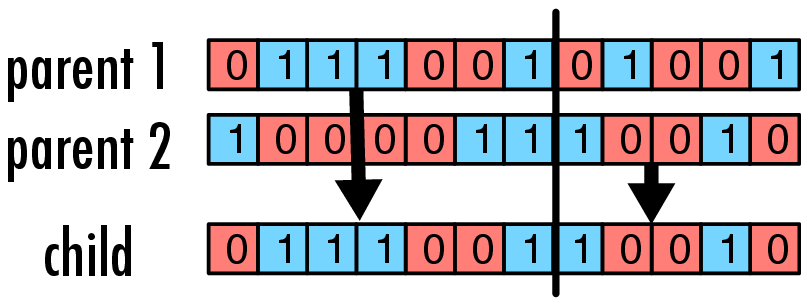}
\includegraphics[width=0.3\textwidth]{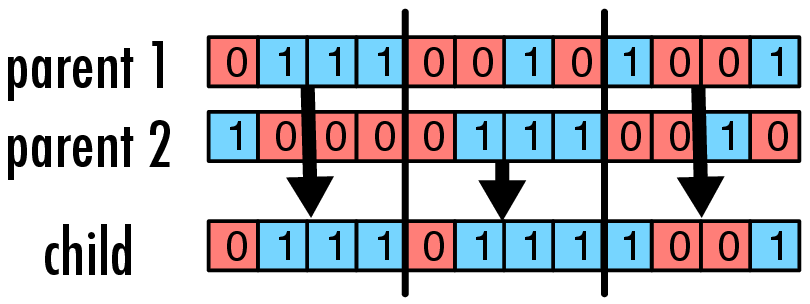}
\includegraphics[width=0.3\textwidth]{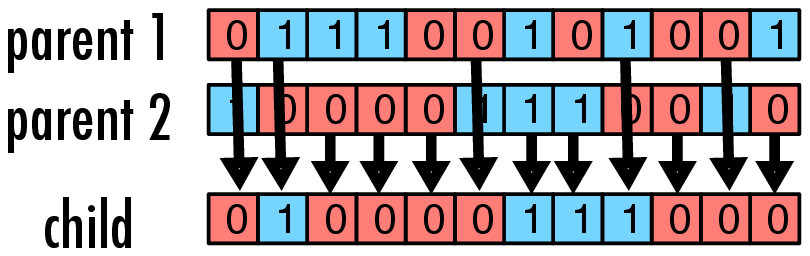}
\caption{Examples of used breeding: cross-over one random point on the left,
cross-over two fixed points in the middle 
and random on the right panels.} 
\label{F:Breed}
\end{figure}

After selecting the parents, we need to produce the new generation, this can be achieved through  
``breeding''. Breeding is the rule according to which a child is produced from the selected parents.
The genes of the child are constructed by mixing the corresponding genes of each parent. 
We take one part from the first parent and the other part from the second one. 
Depending on which parts are chosen, there are several types of breeding.
We usually use three different types: 
cross-over one random point,
cross-over two fixed points,
and random.
For the cross-over one point, we choose one bit (denoted by $i$) randomly
as the cross-over point and the child's genes are
created by combining the first $i$ bits of the genes first parent with the last $N-i$ bits of the genes 
of the second parent (see the left panel of the Figure~\ref{F:Breed}).
For the cross-over two fixed points, the genes of the child are
built from three equal parental parts (see the middle panel of the
Figure~\ref{F:Breed}). 
In the random breeding,  each child's bit is chosen randomly from 
the corresponding bits of the parents (see the right panel of the
Figure~\ref{F:Breed}).


\subsection{Mutation}
\label{SS:Mutation}
The first generation is chosen randomly by drawing parameters
uniformly within the priors specified in~\cite{Babak:2008sn}. The chosen selection
implies that the quality of our organisms is likely to be  increased
with each generation.  But, if we use only these two processes, the
range of resulting genes is quite restricted: it totally depends
on the initial random state and is just a combination of the parts from
the first generation. The combination of genes and therefore the exploration of the 
parameter space is very limited and completely dependent on the initial choice.
This undesirable feature can be cured by introducing mutation.

\begin{figure}[!htb]
\center
\includegraphics[width=0.49\textwidth, clip=true, viewport=0 7 210 100]{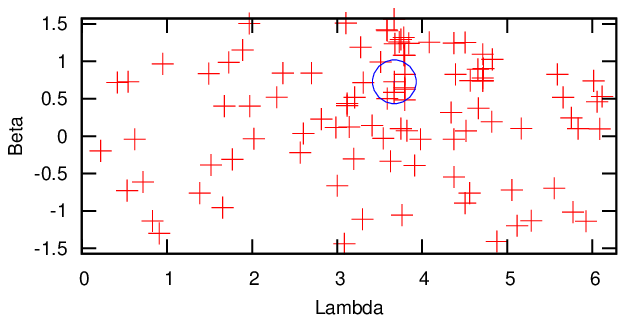}
\includegraphics[width=0.49\textwidth, clip=true, viewport=0 7 210 100]{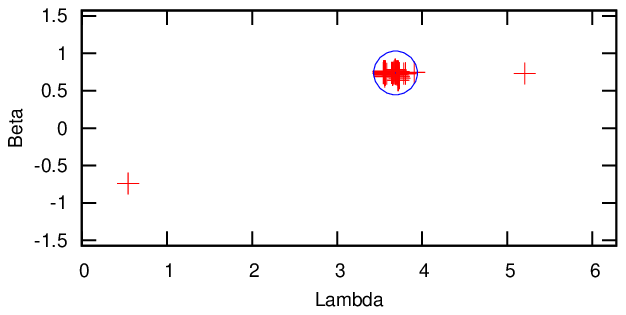}
\caption{Example of distribution of 100 organisms in two dimensions space which is the sky position ($\beta$,$\lambda$). The left panel shows the case of large PMR value (0.5) which corresponds to the beginning of the search. The right-panel shows the case of small PMR value (0.01) which corresponds to the end of the search. The best organism  as well as the true solution is at the center of the blue circle.} 
\label{F:PMRLargeSmall}
\end{figure}

Mutation in GA works in a way similar to how it operates in the nature. 
Mutation is a random change of few alleles in a gene of an organism;
in our algorithm it corresponds to changing few bits in a representation 
of a parameter value of a template. 
The probability of mutation 
  is  called the probability mutation rate (PMR). We
mutate each gene of each child independently and there are several types of
mutation. First we need to decide whether we mutate a gene or not,
and, if yes, we need to decide on the mutation rule (how we do it). 
The first possibility is that we always mutate the gene and mutation
is applied to each bit of gene independently. 
Each bit is flipped with probability PMR. The second possibility is to mutate a gene with probability PMR.
In this case we have used two different
rules to mutate the gene: (i) we flip N randomly chosen bits (ii) we flip N adjacent
bits. Different types of mutations together with the value of PMR
define the exploration area of the parameter space.
An example is shown in  Figure~\ref{F:PMRLargeSmall}, in which we start with  $\textmd{PMR} = 0.5$ 
at the beginning of the search (left panel, one can see  that the templates are scattered all over the space) and then slowly reduce it to $\textmd{PMR} = 0.01$ (the right panel). The true solution is located in the center of the blue circle. We will come back to PMR again in the Section~\ref{SSS:EvolPMR}.


\subsection{Tuning the algorithm using code, breeding and mutation}
\label{SS:Experiment}

In order to get comprehensive understanding of  all kinds of
representation, breeding and mutation to tune our search, we did the 
systematic experiments to test one degree of freedom at the time.
We fixed GA configuration and allow only one parameter to vary and analyzed
the results of the search. We have tested PMR, number of organisms 
in the generation, gene representation (binary, Gray, alternating) type of 
breeding and mutation.  We would need a separate paper to summarize 
this study, it is not our intend here, we will just give a small example below.


 We  found that  
 alternating the binary and Gray representation is more effective than using only 
 one of them. We characterized different types of breeding and mutation according to
 the resulting exploration area of the parameter 
 space. The result of three  such combinations is summarized 
 in the Table~\ref{T:CclBreedMute}. Based on this result we decided 
 to start the evolution with exploration of the large part of the parameter 
 space  (BCO1R-MNR8), then continued with BCO1R-MA, and finally, as our algorithm converged to the 
 solution we explored  small area around the best point intensively (refinment with
 BR-MNA8).
 


\begin{table}[htdp]
\begin{center}
\begin{tabular}{c|c|c|c|c}
    & \multicolumn{2}{c|}{Combination} & \multicolumn{2}{c}{Width of Exploration}  \\
Name & & & \\
&  Breeding & Mutation &  large area &  local area \\
\hline
BCO1R-MNR8 & Cross-over one random point & N bits randomly & Greatly & No \\
BCO1R-MA & Cross-over one random point & Each bits independently & Yes & Yes \\
BR-MNA8 & Random & N adjacent bits & No & Greatly \\
\end{tabular}
\end{center}
\caption{Impact of different types of breeding and mutation on exploration of parameter space.}
\label{T:CclBreedMute}
\end{table}


\section{Acceleration of genetic algorithm}
\label{S:Acc}
We have introduced above three fundamental concepts
used in any GA (selection, breeding and mutation), which might be sufficient for 
a simple search. However, in our
case (multi-dimensional parameter space with many local maxima)
it might require a large number of iterations 
with the possibility that the end results might correspond to a local
maximum. To reduce the required number of iterations and to increase the stability
and efficiency of the algorithm, we introduce  several accelerators
which are used in our search.

\subsection{Standard accelerators}
\label{SS:StdAcc}
In this part, we describe the standard accelerators used in GA.

\subsubsection{Elitism}
\label{SSS:Elitism}
Selection and breeding do not guarantee that the next generation will be better than 
previous one. If we completely replace the old generation with the new one, it is possible 
that  we might lose the organisms with best quality. The overall tendency (trend of the 
evolution) is to increase the quality, but it can go down over some short period of time.

The elitism (or cloning) is a simple way to maintain the quality across generations.
If the best quality of the new generation is lower than the best quality of the current one,
the best organism is propagated to the new generation. It is possible to clone one or several 
best organisms into the new generation. The elitism stabilizes the GA  and guarantees 
the convergence of the algorithm.

\subsubsection{Simulated annealing}
\label{SSS:SimAnn}
The simulated annealing method has been already
employed in LISA data analysis \cite{Cornish:2006ms} and proven to be very useful. 
In this method the smoothness of the quality surface is controlled by the introduced
temperature parameter. If the temperature is high, the quality surface is very smooth and
nearly all the organisms (good and bad) can be selected for
breeding with a similar probability. If the temperature is low, the quality
surface is highly peaked around the maxima and only the best organisms
can be selected. Usually,  a high temperature is selected at the beginning
of the search to have a large area of  exploration. The temperature is decreased as the solution
is approached. 

Temperature is introduced in the selection process through the 
probability of selecting an organism. We set this probability according to  $p_i=q_{i} / \sum_{j}^{N} q_{j}$
where we have the quality of each organism redefined by the introduction  the temperature parameter $T$ as follows:
\begin{eqnarray}
q_i=\exp {\frac{(Q_i - Q_{\textmd{best}})}{T}},
\label{E:RealQ}
\end{eqnarray}
where  $T$ is the temperature, $Q_i$ is the quality  of $i$-th organism
and $Q_{\textmd{best}}$ is the quality of  the best organism. One can see that all $q_i$
are similar if temperature is high.

We devise several kinds of annealing. 
A standard type is the cooling: the temperature
evolves from the initial temperature $T_i$ 
to the final temperature $T_f$ as follows:

\begin{equation}
T(g) = 
\begin{cases}
{T_i \left( \frac{T_f}{T_i} \right)^{\frac{n}{n_c}}} & n < n_c \\
{T_f} & n \ge n_c.
\end{cases}
\label{E:TempCooling}
\end{equation}
where $n$ is a generation number and $n_c$ is the duration of the cooling (in number of generations).  The values
of $T_i$ and $T_f$ are not known a priori. An alternative approach  to control the temperature evolution is to
relate it with the quality of the current generation. The temperature is then evolved according to
\begin{equation}
T= \left( \frac{\rho}{\rho_{th}} \right)^g \quad \textmd{with} \; \rho_{th} = \rho
\quad \textmd{if} \; \rho < \rho_{th},
\label{E:TempSNRle}
\end{equation}
where $\rho = \sqrt{2 Q_{best}}$ (which is the SNR of the best organism if  we use log likelihood as a quality
$Q$) and $g$ and $\rho_{th}$ are two additional parameters. Usually we use $g = 2$  which corresponds to the
thermostated annealing introduced in \cite{Cornish:2006ms}. In the beginning we keep the temperature equal to
unity, and a high PMR is used to explore  the parameter space and build up the SNR. On reaching $\rho_{th}$,
heating
is switched on to increase the exploration area by smoothing the likelihood surface and to force the colony to search 
for a higher maximum. Periods of high temperature are alternated with periods of low temperature (in a periodic manner),
this way the regions around the local maximum and the global parameter space are explored in turn.   

\subsubsection{Evolution of PMR}
\label{SSS:EvolPMR}
 As mentioned above, another way to control the volume of exploration  is by varying the PMR  (see Section~\ref{SS:Mutation}). 
Usually we start with a large value for the PMR
(about 0.2), which is then gradually decreased to give more importance to the
breeding. In the end, the search becomes stationary and approaches
the true solution, so the PMR needs to be quite low (usually we decrease it down to 0.01). 
The typical spread of the organisms in the beginning of the search is depicted in the left panel of the Figure~\ref{F:PMRLargeSmall} and we slowly evolve it towards the right panel by decreasing PMR. 

The three most frequently used types of PMR evolution in our analysis are 
(i) cooling, (ii) fixing and cooling  and (iii) genetic genetic algorithm with PMR.

In the first case of cooling, the PMR evolves from the initial value
$\textmd{PMR}_i$ at generation $n=0$ to the final value $\textmd{PMR}_f$ at generation $n = n_c$ 
according to
\begin{equation}
PMR = 
\begin{cases}
{\textmd{PMR}_f \left( \frac{\textmd{PMR}_i}{\textmd{PMR}_f} \right) ^{\frac{n_c-n}{n_c}}} & n \le n_c \\
{\textmd{PMR}_f} & \textmd{otherwise}.
\end{cases}
\label{E:PMREvolFixCool}
\end{equation}

In the second case of fixing and cooling, at the beginning, the PMR is fixed as $\textmd{PMR} = \textmd{PMR}_i$ for $n_i$ generations, then it is  cooled  to $\textmd{PMR}_f$ in the next $n_c$ generations as follows:
\begin{equation}
\textmd{PMR}= 
\left\{
    \begin{array}{l}
           \textmd{PMR}_i, \quad \textmd{if} \; n<n_i \\
            \textmd{PMR}_f \left( \frac{\textmd{PMR}_i}{\textmd{PMR}_f} \right) ^{\frac{n_c+n_i-n}{n_c}},
            \quad \textmd{if}\; n\ge n_i.
    \end{array} 
\right.
\label{E:PMREvolFixCool2}
\end{equation}

In the last case of genetic genetic algorithm,  the PMR is treated as an
additional parameter of each organism. The PMR parameter evolves (we search for an optimal value) 
by the genetic operations in the specified range $[\textmd{PMR}_{\textmd{min}},\textmd{PMR}_{\textmd{max}}]$. 

We use all the above types of PMR, for each gene we specify its own evolution path. 
Some parameters converge to the true solution faster than other, and some spin related parameters
have multiple solutions. We use the PMR evolution scheme which reflects the convergence of the 
parameter and uniqueness of the solution.


Note that we control the exploration area by using both simulated annealing and PMR. 
Each of these performs somewhat differently. 
 Simulated annealing acts on the quality of the organism and affects the selection procedure
 for breeding, thus it uses  the combination of the initial genes without adding new. 
 On the other hand the PMR changes the 
 structure of each gene and therefore brings in ``new blood'' into the generation (creates new combinations). 
The best result is usually achieved by combining together PMR with simulated annealing.

\subsection{Accelerators specific for MBH search}
\label{SS:OtherAcc}
In this part, we describe the non-standard acceleration processes
introduced by us and which utilize the properties of the signal
and/or of the antenna beam pattern.

\subsubsection{Brother}
\label{SSS:Brother}
The source sky position is encoded in our model of the signal in the phase
(Doppler modulation) and in the amplitude through the antenna pattern function.
For low frequencies the Doppler term is weak and majority of the information is stored
in the directional sensitivity of the detector. However the antenna pattern function
given in expressions  \eqref{E:TDILW_Fp} and \eqref{E:TDILW_Fc} is symmetric 
with respect to the transformation  $\beta\rightarrow-\beta, \quad \lambda \rightarrow \lambda+\pi$
(mirrored/antipodal sky position). This implies a possible degeneracy in the parameter space, and, indeed 
we observe a high value of the quality at the antipodal sky position, making it very difficult to distinguish between those two. 


In order to overcome this problem, we introduced what we call the brother of the clone. 
With each clone we
associate one organism (brother) created by 
copying the parameters values from the clone and then changing a few
of these value by following particular rules.
In our application of the GA for black hole binaries, the brother  explores
the parameter space around the mirrored (antipodal) sky position of each
clone. In a particular search, the best organism usually jumps between these two sky positions
until it settles on the best solution in terms of the quality.

\subsubsection{Local mutation}
\label{SS:LocalMutation}
What benefit one can have from using binary and Gray representations of the
same parameter?  The reason lies in representation of two adjacent integers in
the binary representation. Two close decimal values of $\theta_{k}$ 
which differ only by $\Delta_{x}$ (i.e. corresponding integers differ by 1), may differ 
by several bits in their binary code.   For
example, in the standard binary representation, the separation between the gene 
value 011111 and 100000 is equal to the resolution $\Delta x$ (i.e. minimal distance), but, as
one can see, it is necessary to flip all the bits for making this small change in the parameter value. 
This problem can be solved in two ways: (i) by alternating the Gray representation (where two adjacent 
integers differ by one bit) with binary, and (i)  by introducing the ``local
mutation''.  Local mutation is a small (of order few $\Delta x$) random
change in the parameter value which can push it across the boundary. 
Note that the binary and Gray codes have different bit boundaries, so the alternation 
between them helps in the global exploration of the parameter space, whereas  
the local mutation  helps the organisms to cross a particular bit boundary
and works  locally.

\subsubsection{Fixing the significant bits}
\label{SS:FixBits}

During the test runs of the GA, we  noticed that some parameters
are very well estimated already after few hundred generations.
For example, the time of coalescence $t_c$ can be 
found with high precision in less than 200 generations. 
By restricting the search range of these well estimated parameters, the search efficiency can be
improved. We achieve this by fixing (freezing) the most significant bits of such
parameters which reduces the allowed dynamical range. This significantly speeds up the 
search. Note that we might still keep the PMR  for this gene high in order to have an efficient
exploration of the restricted parameter space.

\begin{figure}[!htb]
\center
\includegraphics[width=0.7\textwidth]{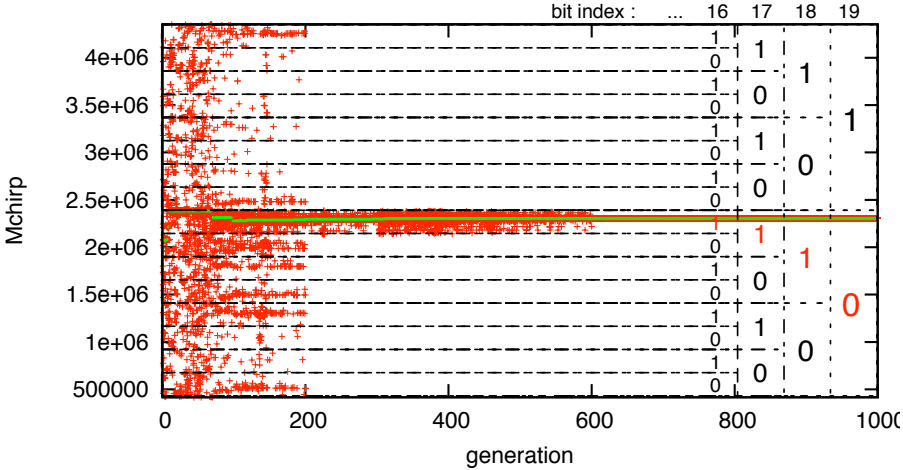}
\caption{Example of the chirp mass exploration by the colony of organisms.
 The green points correspond to the position of the best organism. The separations shows the structure of the binary representation.  The numbers on
 the right are values at bit positions listed on top.} 
\label{F:Ex_FixBits}
\end{figure}

Let us give an example how it works in practice. The Figure~\ref{F:Ex_FixBits} shows
a typical example of the chirp mass, $M_{c}$, exploration in our search. 
This parameter is encoded using 20 bits. First 200 generations have no restriction 
and a large PMR is used so the colony explores the whole  of the prior range. 
However  the higher concentration of the organisms around the best one
(depicted by a green line)  can be noticed which reflects its high quality and, therefore
 proximity to the true solution.
After the $200^{\textmd{th}}$ generation  we fix the bits at a position higher than a  randomly chosen 
number between $14$ and  $16$. It means that the bits $b_{Mc}[16]$, $b_{Mc}[17]$, $b_{Mc}[18]$, $b_{Mc}[19]$ (and sometimes $b_{Mc}[14]$, $b_{Mc}[15]$) of all the organisms are fixed to the value of the best organism (1,1,1 and 0 here). It shrinks the search area to $[2138483.938, 2384509.746] M_{\odot}$ which corresponds to
\begin{eqnarray*}
\textmd{lower boundary} & = & M_{c,min} + \Delta M_{c} \left( 0 \times 2^{19} + 1 \times 2^{18}  + 1 \times 2^{17} + 1 \times 2^{16} \right) \\
\textmd{upper boundary} & = & M_{c,min} + \Delta M_{c} \left( 0 \times 2^{19} + 1 \times 2^{18}  + 1 \times 2^{17} + 1 \times 2^{16} + \left(2^{16} -1 \right) \right)
\end{eqnarray*}
After the $600^{\textmd{th}}$ generation we try to restrict the range further by fixing all the bits
starting at the position $8^{\textmd{th}}$ or $9^{\textmd{th}}$ (again randomly chosen), which corresponds
to narrowing down the range $\Delta M_{c} \times 2^{9} = 1922.106 \sim 2000 M_{\odot}$.
Note that, we can still release  the bits (or change the random range) during the evolution to check the robustness of the found solution.

\subsubsection{Specific breeding and mutation}
\label{SSS:SpecBreedMut}
As mentioned above  in Sec. (\ref{SS:Breeding} and \ref{SS:Mutation}),
different types of breeding and mutation have
different properties (main difference is in the exploration area
around the best organism). The genes (i.e. parameters) do not have
the same rate of evolution during the search. For example, the time of
coalescence and the chirp mass converge to their true values quicker than
other parameters. We customize the evolution of each gene by fixing the significant bits
in a similar manner to the example discussed in the previous section. We also alter the 
type of breeding and mutation of each gene, forcing the exploration range to be large at the 
beginning of the search and changing to the types which are more suitable for more intensive local
exploration close to the end.



\subsubsection{Change of environment}
\label{SSS:ChangeEnv}

\begin{figure}[!htb]
\center
\includegraphics[width=0.5\textwidth]{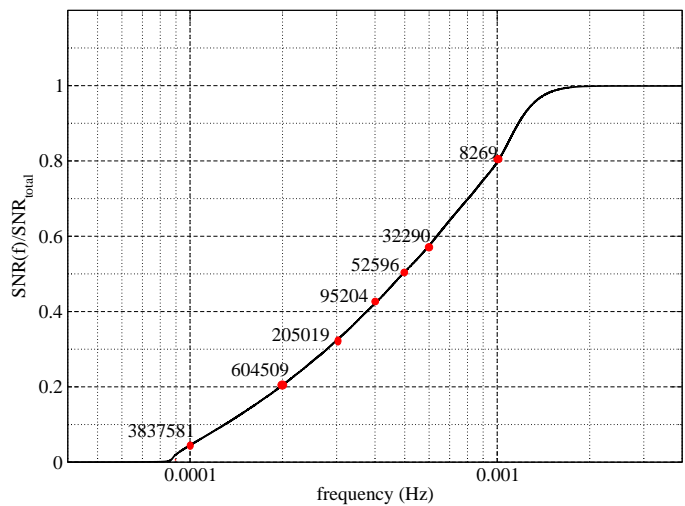}
\caption{Example of accumulation of SNR (MBH-1) as a function of frequency. The points with attached 
numbers show the time until the coalescence in seconds.} 
\label{F:AccSNR}
\end{figure}

While mapping the log-likelihood we have noticed that the binaries that coalesce within the observational time
have more local maxima than the binaries coalescing outside the observational time (this also was mentioned in
\cite{Babak:2008rb} for non-spinning  BHs). 
This can be explained by the accumulation of SNR. Due to the shape of the LISA's sensitivity and
the evolution of the signal's amplitude, the largest part of SNR comes from the last month 
of inspiral. In the Figure~\ref{F:AccSNR} we give accumulation of SNR (scaled by the total SNR) for 
one of the signals analyzed in MLDC3.2. 
We plot the SNR as a function of frequency, we also show (numbers attached to the circles) the time
left to coalescence. As one can see 60\% of SNR comes from the last day and a half of inspiral. 
The above implies that we need to fit only the last day of the signal in order to get a large SNR (in case where we see the coalescence). This is obviously can be done in many ways and this results in multiple maxima of the likelihood.
If the coalescence is not observed 
then we need to fit a large number 
of cycles to accumulate the appreciable SNR, this is harder to achieve unless we are close to the 
correct solution. In this sense it is easier to find the weaker signal with the time of coalescence outside the 
 observational time. We  have utilized this fact in introducing the $A$-statistic which
enhances the low-frequency part of the signal. 
 
 We have also implemented the accelerator which we call ``change of environment''. We put the colony
in different environments and expect the fitter organisms to survive 
in a variety of circumstances. In practice we terminate the template earlier in frequency
and evolve the colony for some time with chopped templates. By changing the frequency range we change
the likelihood surface, the secondary maxima change the size and position, but the global maximum remains
at the same position (location of the true parameters, as shown in the  Figure~\ref{F:DistFMcEtaCutFull}). 
We use this property to alternate between different environments. It helps to move the search away from the local 
maxima where it has a tendency to get stuck, and guides the best organism to the true solution.
It forces the search to seek a better choice of parameters and can also be used to check for the convergence 
of the algorithm to the global maximum.

A typical scheme used in our search is as follows: we start off with a full template and use the Maximized
Likelihood,  $\mathcal{F}_{full}$, as the quality, then we alternate the evolution between full and chopped
templates (``change of environment'') still using $Q = \mathcal{F}$. We finish the evolution of colony using
$A$-statistic. 

We should mention that the frequency annealing introduced in \cite{Cornish:2006ms} helps not only to speed up the
search, but also assists in moving between local  maxima. The structure of the likelihood surface changes as the
duration of the signal increases.

\section{Multimodal search}
\label{S:Multimode}

In this Section we explain how we modify the GA to explore 
multi-modality of the likelihood surface.
As discussed in the  previous sections,
the quality surface have many local maxima. 
Several techniques (simulated annealing, PMR evolution, change environment, etc.)
introduced above, help in finding the global maximum,
but they all assume a single solution and,
therefore, cannot help if there are several maxima of almost equal heights/amplitudes.



\begin{figure}[!htb]
\center
\includegraphics[width=0.49\textwidth, clip=true, viewport=10 5 226 85 ]{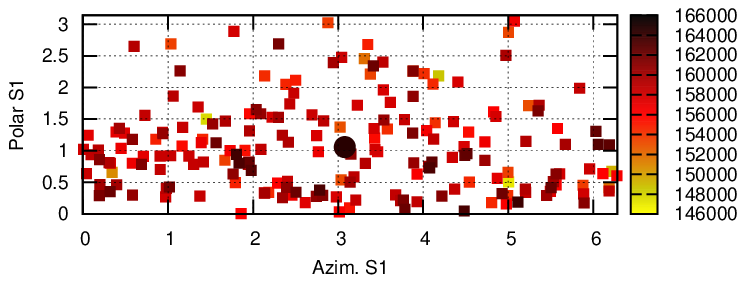}
\includegraphics[width=0.49\textwidth, clip=true, viewport=10 5 226 85 ]{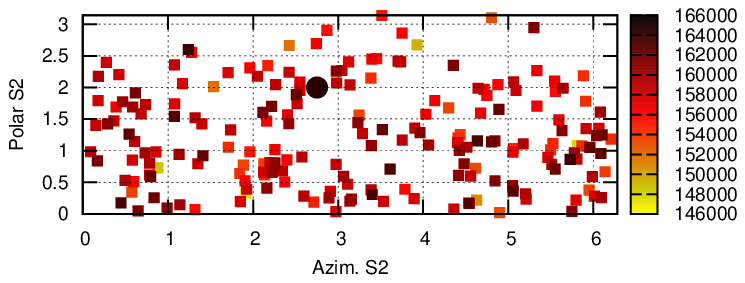}\\
\center 
\includegraphics[width=0.49\textwidth, clip=true, viewport=10 5 226 85 ]{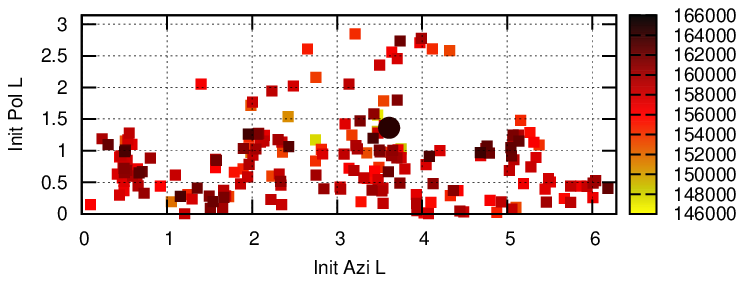}
\caption{Example of the distribution of the best organisms from 196 runs of GA applied to the search for source MBH-3 of MLDC 3.2 (the third signal). The left upper panel shows the initial direction of spin 1, 
the initial direction of spin 2  is in the right upper panel and we plot the initial direction of the 
orbital angular momentum  in the bottom panel. The color scale corresponds to the value of A-statisitc.} 
\label{F:MaximaOfDirLS1S2}
\end{figure}

Five spin-unrelated parameters (time of coalescence,  chirp mass, mass ratio and  sky position) can be 
estimated  using the GA implementation described in previous sections with very
high accuracy. The magnitudes of spins can be also determined in some cases quite well. However other parameters
corresponding to the initial orientation   of spins and of the orbital angular momentum are quite problematic. A
typical situation is presented in the Figure~\ref{F:MaximaOfDirLS1S2}.  We color-code the quality corresponding
to each initial orientation of vectors, it varies  within 12\% of maximum while the points are scattered over the
whole range.  One can see several solutions which are very close in quality to the true one (depicted
by a circle).  The search for a single  maximum will miss other peaks. Instead we want to explore all of them
and, based on the  likelihood of each peak, we can make a claim about possible multiple solutions.


The reason for such a degeneracy lies in the nature of the waveform itself.  
First of all these parameters are highly correlated, 
 second, they enter the expression for the GW phase at higher 
 post-Newtonian orders, and 
 affect the phase and amplitude rather weakly. The later can also explain that 
 we can determine the spins better if we observe the end of inspiral,
 where the contribution from the high order terms is appreciable. 




Another, and most natural, reason for multi-modality  of the likelihood 
is the presence of multiple signals in the data. In the analyzed data set
there were between 4 and 6 signals, but exact number was not disclosed.
The signals usually have different SNR, the search converges to the 
signal with the largest SNR and explores the modes of this signal, other 
signals appear at the initial stage of the search (up to the point at which 
accumulated SNR of different signals is comparable). The main hint
that we are looking at the multiple signals is different values for $t_c$ and $M_c$: 
parameters which are determined most accurately.
 The strongest signal can be removed from the data to recover weaker ones. 
 It is desirable at the end to refine all the parameters 
by using a super-template formed by combining of several signals.

We want to define the mode associated with each local maximum
and explore the parameter space in its vicinity.
The basic idea of our Multimodal Genetic Algorithm (MGA)
is to put a cluster of organism in each mode, to do so we use several clones.
Each clone corresponds to a mode and 
all modes should have comparably high quality. We also increase the size of the colony
so that we keep the number of organisms per clone constant.
The clones participate in the breeding often and attract other organisms and consequently
exploring the neighborhood of each mode. We describe the implementation
 of the evolution later.
 

The crucial point of the MGA is the choice of the clones,
two conditions are necessary for an organism to become a clone. 
First, it should have a quality higher than a certain level. 
This level can be fixed arbitrary or defined relative to the best organism
(for example $Q_{clone} \geq Q^{level} = 0.8 \times Q_{best} $). 
The second condition is that there should not be another clone
on the same mode. 
For that, we define boundaries around a mode 
(i.e. the  rule to separate the different modes)
using the variances $\sigma_{k}^{2} \left( \widehat{\theta}_{clone} \right)$ 
of each parameter $\theta_{k}$ at the clone position $\widehat{\theta}_{clone}$.  
These variances correspond to the diagonal terms 
of the inverse  Fisher Information Matrix (FIM) defined as
\begin{equation}
\Gamma_{i j} = \left\langle {\partial h \over \partial \theta^{i}} \right| \left.  {\partial h \over \partial \theta^{j}} \right\rangle .\label{E:FIM}
\end{equation}
We refer readers to \cite{Vallisneri2008} for details on FIM. 

For each generation we are choose all organisms with $Q_{i} \geq Q^{level}$
as candidates to be cloned. Among these we select only the ones which form the
independent modes:

\begin{eqnarray}
\left| \theta_{k,i} - \theta_{k,clone_{j}} \right| > F_{\theta_{k}} \sqrt{\sigma_{k}^{2}(\widehat{\theta}_{clone_{j}})},
\label{E:ModeSeparation}
\end{eqnarray}
where index $i$ refers to the candidates to be cloned, $j$ to the selected clone and $F_{\theta_{k}}$ is
a factor to control how large should be the distance between the two modes along the parameter $\theta_{k}$.  We
choose $F_{\theta_{k}}$ individually for each parameter and it varies between 15 and 50. This way we define the volume of
each mode.

There are two ways to evolve a colony with multiple clones. The first one is mentioned above,
where we increase the number of organisms proportional to the number of selected clones (modes)
and evolve the system using the GA described in previous sections. If qualities of modes are comparable
we expect to have a fraction of organisms in the close vicinity of each clone, while the remaining organisms explore
 the space in between the modes. Once we have started the evolution, we 
keep the number of clones fixed. If another independent mode is found and its quality is higher 
than the lowest quality among the clones, than the weakest clone (lowest quality) is moved to the 
new found location. Note that we always attach a brother to each clone.



The MGA described above requires a large number of organisms (we need to use at least 10 
organisms per clone). This requires a specific implementation if we want to use a computer 
cluster. We use this algorithm but with a small (less than 10) number  of clones.

The second approach, which we used the most, disallows continuous communication between the modes. 
We perform several independents runs (evolutions) with a single clone. Then we analyze the end results 
of all runs and identify independent modes among them. We use these modes as clones for the next set of independent runs (evolutions). We iterate this procedure until no new modes are found. 
In this approach the modes exchange information discretely, after each single run. 
This multi-runs MGA is described in detail below in Section \ref{SS:MultiStepMGA}.  

\section{Pipeline}
\label{S:Pipeline}

In this section  we describe the chain of algorithms used to arrive at the 
final result presented in the following section.


\subsection{Pre-analysis by a time frequency method}
\label{SS:PreAnalysisTF}
\begin{figure}[!htb]
\center
\includegraphics[width=0.8\textwidth, clip=true]{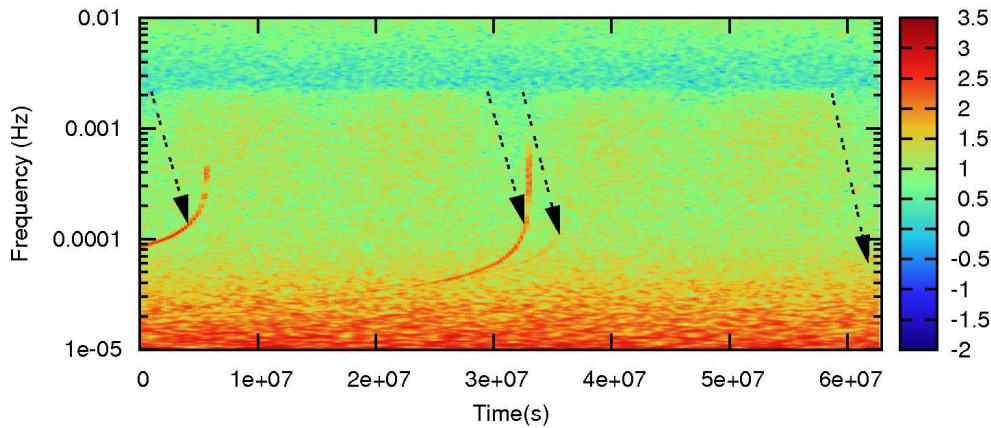}
\caption{Time-frequency representation of channel TDI A of MLDC 3.2. We plot the norm of the  Morlet wavelets transformation.} 
\label{F:PreAnalTF}
\end{figure}

The GW signals from the MBH binaries are usually very strong and do not need very 
sophisticated methods to detect them, especially if we observe the end part 
of the binary evolution. However, it is more complicated if we observe only 
early part of the inspiral. Before analyzing  the data with GA we looked at the 
 time-frequency map of the data constructed using the Morlet wavelet transform
 (see Figure~\ref{F:PreAnalTF}).
From this map, we can clearly identify three strong signals
with the time of coalescence within the observational time and one weak signal
with time of coalescence about 3 months after the end of observations (signals are pointed by arrows).
As we have mentioned earlier and will discuss in detail later, there is one more 
weak signal which coalesces even later. This signal is low frequency and too weak to be 
seen by eye in the data.

In producing submitted MLDC results, we split the data in three parts,  based on the time-frequency analysis
discussed in the previous paragraph. The first part contains the strongest signal completely and low frequency
parts (few first months) of all other signals. The second part contains two other coalescing binaries and parts
of remaining weak signals. Finally the third part contains only the signals coalescing outside the observational
time. An iterative approach could also have been employed where the strongest signal is found and then removed
from the data which is then analysed to detect other signals. This process is repeated until no more signals can
be found. Estimating the residuals after subtracting the detected signals presents a particular problem in this
incremental approach. A disadvantage of our chosen approach is the lose of some SNR, but we can be sure of
avoiding the corruption of the weak signals by residuals of strong ones. However, it turns out that in order to
find the fifth (the weakest) signal we had to remove the fourth signal (the right most one in
Figure~\ref{F:PreAnalTF}) due to the strong interference with the secondary maxima.

\subsection{Multiple steps MGA search}
\label{SS:MultiStepMGA}

We should not forget that GA is a stochastic search method. We can be sure about the  convergence to the true
solution if the likelihood surface is smooth and uni-modal. Unfortunately it is not the case, we have implemented
many tricks to get through the forest of the local maxima to find the highest peaks. As mentioned earlier, our
algorithm found several solutions with similar values of the likelihood. The evolution can still end up on one or
another of these maxima, depending on the initial state and the seed of the sequence of random numbers. This is
the reason behind implementing MGA. 
We have briefly introduced the multi-runs MGA  in Section~\ref{S:Multimode}, here we give a bit more detailed description. 

In this implementation of MGA, the modes exchange information discretely. We start with $N_{run}^{std}$ runs of a
``single clone + brother + 20 organism''  evolution. We use all accelerators introduced in Section~\ref{S:Acc}.
We call each of these runs ``standard'' (as opposed to the global MGA run/search). In the first step we explore
the parameter space trying to find as many maxima as possible. We evolve each colony for 2500 generations,
 then we collect the results of all these evolutions and identify the modes associated
with the best organisms as described in the Section~\ref{S:Multimode}. In our search we followed 50 best modes.
We attach a colony and start another standard run for each mode, in other words we start $N_{run}^{mode} = 50$
independent evolutions for a single clone plus colony. In this step each mode is either refined or migrates to
a new location outside its boundary, with a higher quality. In addition we restart $N_{run}^{std}$
standard evolutions searching for more  modes. At the end of this step, the results from the $N_{run}^{mode} +
N_{run}^{std}$ runs are  collected and  a new set of modes is identified. We iterate the process until the 50
best modes  do not change anymore. 

We found that number of strong modes depends on the parameters of the signal and therefore keeping the number of
modes to be explored fixed is unreasonable. In the post-MLDC exploration we used a variable number of modes:
keeping  all the modes with the quality within 2\% of the best one.

Both standard and mode exploration runs have similar evolution and differ mainly by an initial state of the 
evolution. We always use simulated annealing and the temperature alternates between hot and cold phases, the 
threshold $\rho_{th}$ (see part~\ref{SSS:SimAnn}), which regulates the temperature,  decreases with the number of
generation for both phases. The number of organisms in each run is kept fixed at 20.
For the standard run, the evolution of the PMR and of the type of breeding and 
mutation are chosen such that the exploration volume is high at the beginning of the search and low towards the
end (i.e. the volume near the vicinity of the best organism is search more intensively).
As evolution progresses we gradually fix the range of parameters in the following order:
$t_{c}$, $M_{c}$,  $\eta$, sky position, amplitude of spins, and finally, the initial direction of the orbital angular momentum and spins.



As mentioned above there is a significant difference in likelihood maps for the signals coalescing within
and outside the observational time. This is reflected in the search strategy for those two types: 
due to fewer number of secondary maxima for signals with coalescence outside observational time,
we used only moderate simulated annealing and change of environment and fixed the ranges of $M_{c}, t_c$
much later in the evolution.



After some iterations the modes reach the stable state: we do not see any new modes and the existing modes are
settled at stationary positions (maxima). At this point we stop the run, and all the modes found constitute our
solution.
One can use a Bayesian approach to assign a probability to each mode by calculating the Bayesian evidence.


\section{Results}
\label{S:Results} 

We have previously shown \cite{GAspinbbh_Proceeding} that GA works very well in the case  of the non-spinning MBH
even without using the multimodal search. In this section we discuss the results of the search for spinning MBH
binaries. We present here the outcome of the  analysis of MLDC3.2. By the deadline we did not implemented the MGA
in full and therefore  we have below two subsections: In the subsection~\ref{SS:MLDC} we give the results
submitted  by the deadline, and in the subsection~\ref{SS:PostMLDC} we present the  results of the  full scale
MGA analysis (obtained after the deadline). The main difference is in the number of recovered modes and switching
to the full LISA response at the end of the search to reduce the bias due to mismatch between response function
used in the signal generation and the one used in our analysis.



\subsection{MLDC results}
\label{SS:MLDC}

The signals present in the data can be split in two types: the binaries with the time of coalescence inside the
observational time and others whose coalescence happened outside the observations. The difference between these
two types is in the number of local maxima, SNR and consequently in the accuracy of the recovered parameters.


\subsubsection{Coalescence within the observational time}
\label{SSS:CoalInObsTime}
We have found three signals of this kind in MLDC3.2.  
In the MGA we restricted the search to only 50 best modes selected at each step.  Among 50 explored modes for each
signal, we have identified a small number of distinctly strong and comparable modes for the submission. After
14th, 8th and 7th iterations respectively of the multi-runs MGA search, we obtained five modes for the strongest
signal with the shortest $t_{c}$ (srcMC1 which is MBH-1 in MLDC notation), four modes for the second one (srcMC2
or MBH-3 in MLDC) and six modes for the weakest  signal (srcMC3 or MBH-4 in MLDC). 


The results are presented in the first half (first three rows)  of the Table~\ref{T:ResError} which lists the
relative/absolute errors, global overlaps and quality for modes submitted in MLDC 3.2 for each signal (without
the direction of the spins and of the orbital angular momentum). These errors should be compared to the
corresponding predictions from FIM which are also given in the Table~\ref{T:ResError} in the row labelled as
``True''. For the chirp mass, the errors for all the modes are similar to the ones estimated from the FIM.  For
others parameters, the errors are generally few times higher than predicted by the FIM. At least part of this
discrepancy  comes from the bias caused by the signal approximation -- we have used  the long wavelength limit 
which is valid for the low frequency part and breaks down near the coalescence. The mode with the error for the
sky position higher than 175 degrees corresponds to the antipodal location on the sky. Taking this as a genuine
degeneracy, we see that the source location is found with the precision higher than 10 degrees for srcMC1,  5
degrees for srcMC2 and one degree for srcMC3.


\begin{table} 
\begin{center}
\begin{tabular}{lr|rrrrrrr|rr}
\hline
Source & mode & $\Delta M_{c}/ M_{c}  $& $\Delta \eta/ \eta $ & $ \Delta t_{c} $ &  $ \Delta $ Sky  & $ \Delta a_{1} $ & $ \Delta a_{2}  $ &  $\Delta D / D$ & $\mathcal{O}$ & quality \\
 & & $\times 10^{-5}$ & $\times 10^{-4}$ & (sec) & (deg) & $\times 10^{-3}$ & $\times 10^{-3}$ & $\times10^{-3}$ & &  \\

\hline
               & True &  1.3  &     4.4 &       6.1 &      1.18 &  2.7  &   6.2  &   6.2 & 1.0 &  392171 \\
\cline{2-11}
               & 1      &  4.3  &  39.8 &       7.2 &      4.15 &  8.6  & 83.0 &    8.7 & 0.99189 &  392628 \\
srcMC1 & 2      &  7.8  &  58.6 &  631.4 &  177.54 &  4.7  & 64.0 &    0.6 & 0.99236 &  392595 \\
MBH-1  & 3      &  2.7 &   15.1 &       0.7 &      5.39 &   5.2  & 84.9  &   3.8 & 0.99198 &  392589 \\
                & 4     &  0.2 &   62.5 &    33.7 &      1.43 &   1.9  & 87.1  & 14.2 & 0.99174 &  392533\\
                & 5     &  2.4 &     6.1 &    62.9 &    11.67 &   7.6  & 47.5  & 80.4 & 0.99235 &  392385\\

\hline
               & True &  4.3 &      7.2 &       9.1 &      0.82  &     2.9  &   5.3 &    7.2   & 1.0 & 164559 \\
\cline{2-11}
srcMC2 & 1     &  8.9 &      5.1 &   100.8 &  175.94 &     6.2  & 18.7 &  27.2  & 0.98965 & 164626 \\
MBH-3  & 2     &  6.4 &  106.6 &  164.4 &   178.49 &  46.2  &  31.8 &    9.4  & 0.99800 & 164608 \\
               & 3     &  1.4 &  106.6 &    39.0 &       3.65 &   42.4  &  39.9 & 27.9  & 0.99592 &  164589\\
               & 4     &  0.5 &  113.9  & 179.9 &  176.53 &   45.1  &  11.1 & 20.5  & 0.99754 &  164583\\

\hline
               & True & 10.1  &      55.3 &    26.7 &       0.47 &    29.2 & 151.4  &   138.7  & 1.0 & 5823.92 \\
\cline{2-11}
               & 1     & 22.0  &    126.2 &  362.2 & 179.82  &    57.4  &    93.1  &  337.9  & 0.99403 & 5845.22 \\
srcMC3 & 2     &  16.8 &    153.9 &  337.7 & 179.84 &     51.6  &  373.4 &  252.2   & 0.99752 &  5832.83\\
MBH-4   & 3     &    1.1 &    166.8 &    30.7 &     0.19  &     51.9  &  385.2 &  252.2   & 0.99463 &  5832.07\\
                & 4     &  29.2 &   303.4 &  349.5 & 179.79 &     28.8  &  401.7 &  220.8   & 0.99686 &  5832.01\\
                & 5     &    4.5 &     75.2  &    31.4 &      0.12  &    47.1  &  173.7 &    90.8   & 0.99935 &  5832.01\\
                & 6     &  29.2 &  138.5  &  258.9 &  179.27  & 226.4   & 184.0 &  125.3  & 0.99710 &  5830.61\\

\hline
\hline
               & True &    167.3 &  702.8 &   14641.2  &    10.37  &  725.6 &  902.6  & 167.9  & 1.0 &  184.99 \\
\cline{2-11}
srcMC4 & 1     &     251.3 &  778.7 &  18562.2  &    45.07  &    38.3  & 150.6  &    93.3  & 0.936389 &  197.77\\
MBH-2  & 2     &     270.2 &  118.3 &    8405.7  &    10.47 &   218.5  &  251.4 &  178.4  & 0.965423 &  197.31\\
               & 3     &  1114.1 &  952.2 &  38160.8  &  171.10 &  331.7  &  409.0 &  153.0  & 0.943096 &  197.08\\
               & 4     &    714.4 &  1104 &      7942.4  &  141.59 &    11.7  &  665.2 &  169.3  & 0.935997 &  196.00\\

\hline
srcMC5  & True &    315.1 &    670.3 & 73890.8  &     6.40 &  453.7  & 699.0  &  321.4  & 1.0 &  38.75  \\
\cline{2-11}
 MBH-6  & 1     &  1042.3 &  1235.6 & 82343.2 &      2.11 &  258.2  & 191.6  & 260.5  & 0.929130  &  47.41 \\
                & 2     &    293.7 &    618.8 & 43456.8 &  173.94 &    89.6  & 122.9  & 430.5  & 0.729048 &  41.78 \\

\hline
\end{tabular}
\end{center}
\caption{ Relative/absolute errors, global overlap, $\mathcal{O}$, and quality for the modes submitted in MLDC~3.2.  All parameters are defined in~\ref{SS:Model}. $\Delta \textmd{Sky}$ is the angular distance in the sky between the true and the estimated positions. The second column gives the mode number. The errors for true parameters are obtained using the FIM. For the tree first sources (srcMC1, srcMC2 and srcMC3) which coalesce during the observational time the quality corresponds to $A$-statistic and the two others (srcMC4 and srcMC5) which coalesce after the end of observation the quality is the Maximized Likelihood $\mathcal{F}$.}
\label{T:ResError}
\end{table}

We found a strong degeneracy in the initial directions of the orbital angular momentum and spins, so we decided
to submit several well separated modes. Only for srcMC2, one of these modes corresponds to the true 
parameter set. For srcMC1 and srcMC3, the true mode was missed, however we found it in the full scale 
MGA analysis conducted after the deadline (see subsection~\ref{SS:PostMLDC}).

The last two columns of the Table~\ref{T:ResError} show the value of $A$-statistic (quality column) for each mode and the multi-stream overlap defined as 
\begin{equation}
\mathcal{O}  (  \widehat{\theta}_{e} ) =  \frac{ \langle h_{A} (  \widehat{\theta}_{e} ) \mid h_{A} (  \widehat{\theta}_{t} )  \rangle + \langle h_{E} (  \widehat{\theta}_{e} ) \mid h_{E} (  \widehat{\theta}_{t} )  \rangle }{ \mathcal{N}[h(\widehat{\theta}_{t})] \; \; \mathcal{N}[h(\widehat{\theta}_{e})] }
\label{E:OverlSNR}
\end{equation}
where $\widehat{\theta}_{t}$ corresponds to the true parameters, $\widehat{\theta}_{e}$ are our estimated parameters and $\mathcal{N}[h(\widehat{\theta})] $ is the norm of the template $h(\widehat{\theta})$ defined as
\begin{equation}
\mathcal{N}[h(\widehat{\theta})]  = \sqrt { \langle h_{A} ( \widehat{\theta} ) \mid h_{A}  ( \widehat{\theta} )  \rangle + \langle h_{E} ( \widehat{\theta} ) \mid h_{E}  ( \widehat{\theta} )  \rangle}.
\label{E:Normh}
\end{equation}
The overlap $\mathcal{O}$ varies between -1 and 1 (from perfect anti-correlation to perfect correlation) 
and it tells us the loss in the SNR = $\mathcal{N}[h(\widehat{\theta_t})]$~\footnote{Here we have defined the
theoretical SNR as an average over the ensemble of the noise realizations.} due to the mismatch between the
signal and the template.
The SNRs of sources srcMC1, srcMC2 and srcMC3 are 1670.58, 847.61 and 160.51 respectively.

All of these modes have an overlap with the true solution higher than
$99\%$.\;
The value of $A$-statistic as well that of Maximized Likelihood for the recovered modes  is higher than the
corresponding values for the true parameter set. This is a manifestation of the mismatch between the signal and
the template and indicates the importance of using the full response towards the end of the search. We were aware of
this but did not have time to implement it completely before the MLDC submission deadline. Nevertheless, given
this bias in the search, our results are still quite accurate.



\subsubsection{Coalescence beyond the observational time}
\label{SSS:CoalOutObsTime}

During the search we found two signals of this kind. The results are presented in the second half (last two
rows)  of the Table~\ref{T:ResError}. Those are low frequency signals, so our long wavelength approximation works
very well resulting in very small or no bias in the parameter estimation due to the mismatch between the response
functions as discussed in the previous section.

First we identified the source with SNR 18.63 (which is srcMC4 or MBH-2 in MLDC notations). For this signal we
found several modes after 8 steps of MGA search, out of which we selected four modes with highest quality for the
submission. From the Table~\ref{T:ResError} it can bee seen that the errors in the spin independent parameters
are similar to the errors predicted by the FIM. Spin-orbital and spin-spin couplings enter the  phase at 1.5 and
2 PN orders respectively, and since we do not observe the end of the inspiral, these terms contribute very little
to the phase as well as to the amplitude modulation (see the orbital  frequency dependent term in equations
\eqref{eq:S1}-\eqref{eq:L}). Therefore the spin related parameters are intrinsically poorly identified for these
sources which is reflected in our results.


The fifth and the last source is the weakest. In fact it was completely contaminated by the secondaries  of the
srcMC4. In order to identify this source we had to remove the fourth signal. We identified the srcMC4 with the
best (highest quality) recovered mode, generated the signal and subtracted it from the  time series. After that
we repeated the search and already the first standard run found the mode with $\textmd{SNR}>7$ which was a
positive detection. Before the deadline we could  perform only 3 steps of the MGA, however this turned out to be
sufficient, as is indicated by the overlap  column in the Table~\ref{T:ResError}. We have clearly identified two
modes with the opposite sky positions. The SNR for this signal was 12.82 and consequently the parameters have
large uncertainties. The initial directions of the spins and the orbital angular momentum could not be identified
at all. Other uncertainties are consistent with the FIM. This, fifth signal, was correctly identified only by us
among all the participants of MLDC3.2 (at least with the precision which gave an overlap of 0.92).
\subsection{Post-MLDC results}
\label{SS:PostMLDC}

After the MLDC submission deadline, we  finalized the implementation of the MGA 
(this time we have kept all the modes within some fraction of the maximum) and have performed the 
search to completion. We have also incorporated the full
LISA response in our template using LISACode ~\cite{LISACode,LISACode_web2} to refine the final solutions. We
discuss the details below.

For the few first steps we kept all the modes with quality higher than 50\% of the best one, then we increased
the mode selection threshold to 90\% (or higher, depending on the number of  modes detected for a given
signal). We also improved the mode separation criteria by adjusting  $F_{\theta_{k}}$ based on the detailed study
of the quality distribution.
Finally, we  have also added two final search steps using the templates with the full TDI response. 
Here we used a lite version of LISACode simulator~\cite{LISACode,LISACode_web2}. The lite version 
contains some fine-tuned trick which allowed us to compute the two-years long template in less than 15 seconds. 
The final steps with the full TDI response are required only for the signals which coalesce within the
observational time. Only those signals propagate to high frequency where the long wavelength approximation is not
accurate any more, and the SNR is high enough this to matter. Including the full response also helped for srcMC1
to promote the mode (increase its quality) closest to the true solution and slightly suppress the others.

\begin{figure}[!htb]
\center
\includegraphics[width=0.6\textwidth, clip=true, viewport=5 0 170 110]{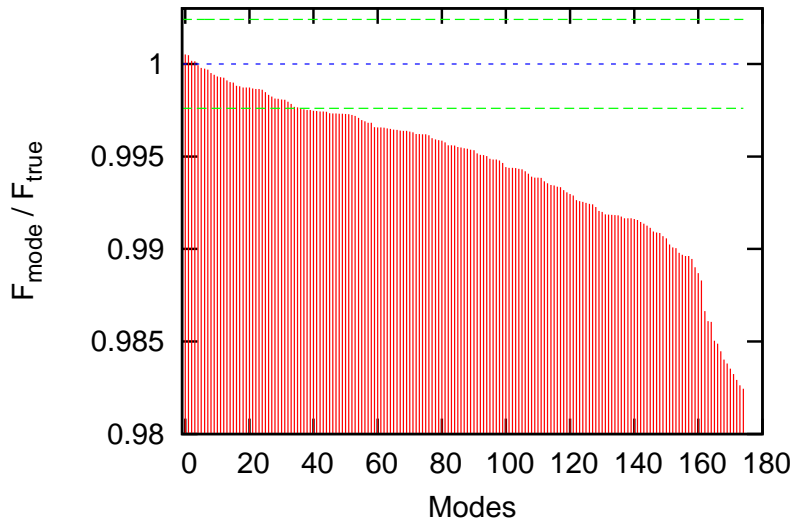}
\caption{Ratio between Maximized likelihood of  modes and true Maximized Likelihood for source srcMC2. 
The blue dotted line corresponds to a ratio equal to one.
The green dashed lines correspond to $\Delta SNR_{i} < 1$}
\label{F:HistosrcMC2}
\end{figure}

For the last, full response search, we selected the modes within 98\% of the best one. This results in the
selection of 26 modes for source srcMC1 after 13 steps of MGA, and 175 and 17 modes for the sources srcMC2 and
srcMC3 respectively after 9 steps of MGA .
Figure~\ref{F:HistosrcMC2} shows the distribution of the Maximized Likelihood with the full LISA response 
relative to the true one, $\mathcal{F}_{Full,i}/\mathcal{F}_{Full,true}$, for the modes of the source srcMC2. 
For this source, 36 modes have SNR \footnote{ Approximating $\textrm{SNR} \sim \sqrt{2 \mathcal{F}}$ we have the
following relation between $\mathcal{F}$  ratio and $\Delta SNR$ :
\begin{equation}
\frac{\mathcal{F}_{i}}{\mathcal{F}_{True}} \approx {\left( 1 -  \frac{\Delta SNR}{SNR_{true}} \right)}^{2}
\label{E:DSNR_FiFtrue} \nonumber
\end{equation}
} within one standard deviation of the SNR$_{true}$: $\Delta \textrm{SNR}_{i} = |\textrm{SNR}_i - \textrm{SNR}_{true}| < 1$, and 
4 of them have $\mathcal{F}_{Full,i}$ higher than for the true waveform. Deviations of order unity in SNR can
be easily produced by noise, note that besides the stationary Gaussian instrumental noise we also had
cyclo-stationary Galactic confusion noise. There are similar results for other sources. For srcMC1 we
identified 21 modes with $\Delta SNR_{i} < 1$ and 6 of them have $\mathcal{F}_{Full,i} >
\mathcal{F}_{Full,True}$; for srcMC3, 21 modes have $\Delta SNR_{i} < 1$ and 6 of them with $\mathcal{F}_{Full,i}
> \mathcal{F}_{Full,True}$. We confirmed these results also with the signals generated using
syntheticLISA~\cite{SyntheticLISA}, the simulator used to produce the data set, to avoid possible error
coming from the use of two different simulators.



The difference between identified modes is within the fluctuation that can be caused by the noise, therefore
there can not be a unique solution. However, all the modes are single valued in the non-spinning parameters and
 they split for the initial directions of the spins and the orbital angular momentum (sometimes also antipodal sky
location). 


Our post-MLDC results are presented in Table~\ref{T:ResPostMLDC} in which we list the parameter estimates for
three modes found for each source. These modes are described as follows:
\begin{enumerate}
 \item `B'-mode (`B'-est) is the the mode with the highest Maximized Likelihood value using the template
 with the full response.
 \item `C'-mode (`C'-losest) is the mode closest to the injected signal in all parameters.
 \item `A'-mode (`A'-strophysically relevant) is the mode with the smallest error in the most relevant parameters from the
 astrophysical point of view (sky position, distance, masses, spin amplitude and time of coalescence).
 \end{enumerate}
We estimate the ``closeness'' to the true parameters as 
$$
d \equiv  \textmd{max}_{k}\left\{ \frac{\sigma^{\theta_k}}{\sigma^{\theta_k}_{FIM}} \right\},$$
where  $\sigma^{\theta_k}$ is an error in the estimation of parameter $\theta_k$ and 
$\sigma^{\theta_k}_{FIM}$ is a corresponding prediction from the FIM. 

 Besides better identification of the modes we have also improved the parameter estimation
 which is reflected in increase of the overlap to $ > 0.999$ compared to 0.99 for the MLDC results.


\begin{table} 
\begin{center}
\begin{tabular}{lr|@{\;\;}r@{\;\;}r@{\;\;}r@{\;\;}r@{\;\;}r@{\;\;}r@{\;\;}r@{\;\;}|r@{\;\;}r@{\;\;}r@{\;\;}r@{\;\;}|r@{\;\;}r@{\;\;}}
\hline
Source & mode & $\Delta M_{c}/ M_{c}  $& $\Delta \eta/ \eta $ & $ \Delta t_{c} $ &  $ \Delta $ Sky  & $ \Delta a_{1} $ & $ \Delta a_{2}  $ &  $\Delta D / D$ 
& $\Delta \widehat{S}_{1}$ & $\Delta \widehat{S}_{2}$ & $\Delta \widehat{L}$ & $\Delta \phi_{c}$
& $\mathcal{O}$ & $\mathcal{F}_{Full}$ \\

& & $\times 10^{-5}$ & $\times 10^{-4}$ & (sec) & (deg) & $\times 10^{-3}$ & $\times 10^{-3}$ & $\times10^{-3}$
& (deg) & (deg) & (deg) & $\times 10^{-2}$
& &  \\

\hline
               & True &  1.3  &     4.4 &       6.1 &      1.18 &  2.7  &   6.2  &   6.2         & 8.30  &    6.06  &   1.07 & 7.5        & 1.0 & 1387732 \\
\cline{2-15}
srcMC1 & A        &   10.1 &    9.2 &   25.7 &   1.92 &   0.7 &   44.3 &   13.1       & 64.39 &   79.59 & 84.02 & 259.6       & 0.999870 & 1387772 \\
MBH-1  & B, C   &   13.5 &    8.6 &   24.6 &   8.04 &   6.7 &   28.9 &   22.4       &   9.39 &   15.70 &   7.70 &    14.7       & 0.999944 &  1387946 \\
               &            &   10.8 &   6.3 &   24.2 &   7.04 &   7.1 &   21.8 &   20.8        & 48.18 &   19.07 &   6.20 &    40.6      & 0.999952 &  1387914 \\

\hline
               & True &  4.3 &      7.2 &       9.1 &      0.82  &     2.9  &   5.3 &    7.2     &     1.52 &      3.29 &     0.95 & 2.9         & 1.0 & 355588 \\
\cline{2-15}
srcMC2 &  A      & 13.6 &    4.8 &    29.0 &      1.33 &     2.8 &   28.3 & 19.8    & 104.65 & 155.92 & 138.55 &     4.1       & 0.999939  & 355755 \\
MBH-3  &  B      & 11.0 &  92.5 &  154.6 & 176.51 &   24.1 &     3.5 & 24.7    & 55.37   &   54.86 &    81.31 & 135.7      & 0.999827 & 355769 \\
               & C      &  15.6 & 44.6 &   158.9 & 169.3 &    52.4 &   15.0 & 66.1     & 16.49  &   64.82 &    14.68 & 25.0         & 0.997845 & 354301 \\

\hline
               & True & 10.1  &      55.3 &    26.7 &       0.47 & 29.2 & 151.4 &   138.7         & 22.90 & 65.30 & 16.17 & 102.2      & 1.0 &  12814.2 \\
\cline{2-15}
srcMC3 &  A, B, C  &  9.2  & 139.0 &   40.5 &      0.34 & 55.9 & 390.1 & 181.3          & 159.55 & 74.93 & 63.38 &     7.3      & 0.999311 & 12834.4 \\
MBH-4  &                 & 17.7 &     8.5 & 234.0 & 179.48 & 96.5 & 506.7 & 319.6          &   60.55 & 87.22 & 42.61 & 413.2      & 0.998723 & 12818.8 \\
\hline
\end{tabular}
\end{center}
\caption{ 
 Post-MLDC results. We present relative/absolute errors, global overlap, $\mathcal{O}$, and Maximized
Likelihood using full response $\mathcal{F}_{Full}$ for selected modes for sources srcMC1, srcMC2 and srcMC3. 
All parameters are defined in~\ref{SS:Model}. $\Delta \textmd{Sky}$, $\Delta \widehat{S}_{1}$, $\Delta
\widehat{S}_{2}$ and $\Delta \widehat{L}$ are the angular (geodesic on a sphere) distance between the true and
the estimated direction for, respectively the sky position, the spin of $\textrm{MBH}_{1}$, the spin of
$\textrm{MBH}_{2}$ and the orbital angular momentum. The selected modes are given in several rows for each source
and marked as: `A' corresponding to the mode closest to the true one in the most relevant astrophysical parameters
(sky position, distance, masses, spin amplitude and time at coalescence); `C' corresponding to the mode closest to
the true one in all parameters; `B' is the mode with the best $\mathcal{F}_{Full}$. The mode without any mark is
the second closest.} 
\label{T:ResPostMLDC}
\end{table}







\section{Summary}
\label{S:Summary}

In this paper we have described the application of the Genetic Algorithm to the the problem of detecting
gravitational wave signals from inspiralling spinning MBH binaries and estimating their parameters. We described
how GA can be translated to the problem of GW data analysis, and introduced some custom-designed accelerators of
the evolution which allow us to efficiently explore the 13-dimentional parameter space. In addition to the
standard $\mathcal{F}$-statistic which is popular in the GW data analysis, we introduced a new detection
statistic called  $A$-statistic, which enhances the low frequency part of the signal. Use of $A$-statistic
allows us to partially compensate for the mismatch between the template and the true signal and to change the
structure of the quality surface eliminating some of the secondary maxima.

We have found that the likelihood surface is highly multimodal with several modes having very high amplitudes. In
order to incorporate this in our search we have extended the standard GA to the Multimodal Genetic Algorithm.
We cluster strong local likelihood maxima in the parameter space within 
the volume defined by the slightly enlarged error boxes predicted by inverse of the Fisher information
matrix.
To each cluster or ``mode'' identified in such a manner, we attach a colony of the organisms. The colonies
explore their local regions intensively and exchange the information after every 2500 generations.

We apply this method for the analysis of MLDC3.2 data set. In the blind search, we have successfully found all 5
signals and the recovered solutions have overlap higher than 99.2\% for the strong (high SNR) signals and higher
than 93\% for the  weak signals. The results submitted by the deadline did not fully reflect the capability of
our search method, as the implementation was not complete. We have completed the search after the deadline by
allowing MGA to reach the stable solution. We have also used the full TDI response during the last two steps of
our post-MLDC analysis. This has allowed us to recover all modes and reduce the bias in the parameter estimation
due to use of the long wavelength approximation in our search template. We have achieved a remarkable accuracy 
in estimating non-spinning parameters, as well as reasonably accurate estimation of the spin magnitudes if binary
coalesces within the observational window. Our method is at least comparable, if not better, to other very 
successful algorithms such as MCMC with parallel tempering and MultiNest \cite{FerozMN_2009}. The success of the MGA 
in case of the inspiralling spinning MBH binaries gives us the confidence that it should  also prove to be
highly efficient in the search for Extreme Mass Ratio Inspirals (EMRIs).

\begin{acknowledgments}
WWork of A.P. and S.B. was partially supported by DFG grant SFB/TR 7 Gravitational Wave Astronomy and DLR
(Deutsches Zentrum fur Luft- und Raumfahrt). Y.S. was supported by MPG within the IMPRS program. F.F. is
supported by a research fellowship from Trinity Hall, Cambridge. The search was performed on the Morgan cluster
at AEI-Golm and on the Atlas cluster at AEI-Hannover.
\end{acknowledgments}

\bibliography{refr}

\end{document}